\documentclass[a4paper,10pt,onecolumn]{article}

\usepackage{amsmath}
\usepackage{amsthm}
\usepackage{amssymb}
\usepackage[T1]{fontenc}
\usepackage{graphicx}
\usepackage{natbib}
\usepackage{color}
\usepackage{subfigure}

\graphicspath{{figure/}}

\theoremstyle{plain}

\theoremstyle{definition}

\newtheorem{exmp}{Example}

\theoremstyle{remark}

\title{\huge{\textbf{Abstraction and control techniques for non-stationary scheduling problems}}}
\author{G. Innocenti and L. Pretini}
\date{}

\begin{document}

\maketitle

\begin{center}
\begin{minipage}[hc]{0.80\textwidth}
	\begin{center}
		\small{
		Dipartimento di Sistemi e Informatica, Universit\'a di Firenze, \\
		via S. Marta 3, 50139 Firenze \smallskip\\
		Centro per lo Studio di Dinamiche Complesse, CSDC,\\
		Universit\'a di Firenze
		}
	\end{center}
	\begin{center}
      \textbf{R.T.2/2009}
    \end{center}
	~\smallskip\\
		\small{
		\emph{Abstract.} The paper faces the problem of scheduling from a new perspective, trying to bridge the gap between classical heuristic approaches and system identification and control strategies. To this aim, a complete mathematical formulation of a general scheduling process is derived, beginning from very broad assumptions. This allows a greater freedom of manipulation and guarantee the resolution of the identification (and control) techniques. Both an adaptive and a switching strategies are presented in relation to the performances of a simple Round Robin algorithm.
		}
\end{minipage}
\end{center}
~\medskip\\

\section{Introduction}
According to the most theoretical point of view, scheduling regards
the problem of deciding how a number of users, competing for a
common resource or service, should access it in order to grant or
maximize a certain performance.
In the literature such a problem has assumed a variety of formulations
in very different fields \cite{bib:book05,bib:book07,bib:book08,bib:book09,bib:book06}.
For example, in telecommunications the usage of a common physical
medium (i.e. a bus) over a network and, in general, the whole queueing
theory can be interpreted in terms of a scheduling problem \cite{bib:book75,bib:book93,bib:book00}.
In the recent years a relevant interest for this subject has been also
shown in the information technology field, in particular referred to
the execution of the tasks by the computer's CPU \cite{bib:book05b,laplante,blaze,oliver}.
Here, an important branch of research is focused on the real time
scheduling \cite{bib:RT91,ssac,mltijl,fusiello,effective,palopoli}, where the policy has to satisfy strict time
constraints.
In the automation area scheduling can be discovered in the switching
control theory in terms of the procedure used to detect the most
suited controller for the actual working condition \cite{libe,giarre,rtc,palopoli}.

Though the basic problem is quite straight, a number of features may
modify the framework producing very different situations.
For instance, in the problem of neatly accessing a common bus a large
set of users is usually assumed and, moreover, their arrivals are
often characterized as stationary random processes \cite{channel,bib:telec99, bib:telec05, bib:telec02}.
Conversely, in informatics and electronics there exists the problem of
running jobs of minor priority along with other critical tasks needed
to ensure the right functioning of the device \cite{bib:robot94,bib:robot05,effective,mmmh,super,multi}.
Hence, a reduced subset of the system's users is completely known
beforehand and the policy has usually to serve these ones under tight
constraints.
In general, each context has a different set of related assumptions.
Hence, the scheduling policy has to be specifically developed on the
base of both the desired performance and the considered framework.

In this paper we deal with a large set of tasks to be processed.
In particular, we assume that their arrivals and requests may be
described only in terms of statistics and that their features may vary
in time.
Hence, the tasks do not admit a stationary random process
representation and, thus, the scheduling policy has to be thought as
an adaptive law in order to keep the pace of the changing context.
Here, we introduce a solution inspired by control theory.
First, we present a procedure to derive a functioning model of the
system under a time variant flow of tasks or processes.
Then, by exploiting adaptive control techniques we provide a policy
aiming to the optimization of a certain performance index under the
problem constraints.

\section{Problem overview}
In the following we pursue the goal of deriving a suitable model about the effects of the scheduler's choices onto the performances.
As expected, the practical implementation affects the system's behaviour, but its response as consequence of a certain policy deeply depends on the requirements and the flow of the processes.
Then, as a matter of fact, the law between the choice's strategy and the scheduler's performances intrinsically is time varying.
However, in certain situations the sequence of the jobs submitted to the scheduler can be a priori inquired and, thus, the system's dependency from this flow can be precisely modelled \cite{blaze,handbook}.
In other common scenarios, instead, such a sequence is regarded as a stationary stochastic process, whose features are somehow known.
In this case, standard theory and analysis usually assume a fixed scenario, such as the worst or the mean case, and then develop a rather simple policy in order to maximize the chosen performance's index or to satisfy a certain set of constraints \cite{blaze,handbook,laplante,deadline}.
Here, the main concerns just regard both the practical realization of the scheduling algorithm and the adherence of the processes' features to the assumed hypothesis.

In the following we consider a more general scenario.
We assume that the tasks' flow can not be forecast nor modelled as a stationary process.
Hence, the system's behaviour is expected to be time varying, requiring a suited strategy to keep track of its evolution, in order to accordingly tune the scheduler's policy.
To this aim an adequate level of abstraction in needed.
First, we point out a number of quantities, which turn out informative about the scheduler's performance and behaviour.
Then, by means of standard identification methods, we derive a dynamical model describing the influence of the strategy onto the scheduling performance.
In particular, since this relationship depends on the processes entering the scheduler, the resulting system can be parametrized in terms of the jobs' global context.
Moreover, under rather general assumptions, we will show that such parametrized model can be used to formulate a suitable control law optimizing the performance index and that the corresponding controller has a direct expression in terms of choice's policy.

\section{Scheduling scheme}
In order to approach the problem from a general point of view, regardless to the specific implementation, we find useful to set it up in a schematic way.
Even if this simplified version could not represent any possible alternative of modern scheduling routines, we will show that it turns out straightforward to add more complex structures, then closing up to real situations.
Here, for the sake of generality we will just denote the scheduler's users as ``processes'', though they may assume different names as tasks or jobs in different contexts.

Let us consider a single Server and a set of processes which ask at different instants in time for services offered by the former.
Also suppose that the Server is able to grant its services only to one process at a time and assume that the jobs asked for by the processes could have to be ``interrupted'' until a certain constraint is satisfied.
The choice of the process to serve influences directly the performances and it represents the research's goal.\\
From this basic idea, some logic constructs to define the scheduling algorithm are in order.
\begin{figure}[htb]
\centering
\includegraphics[width=0.8\textwidth]{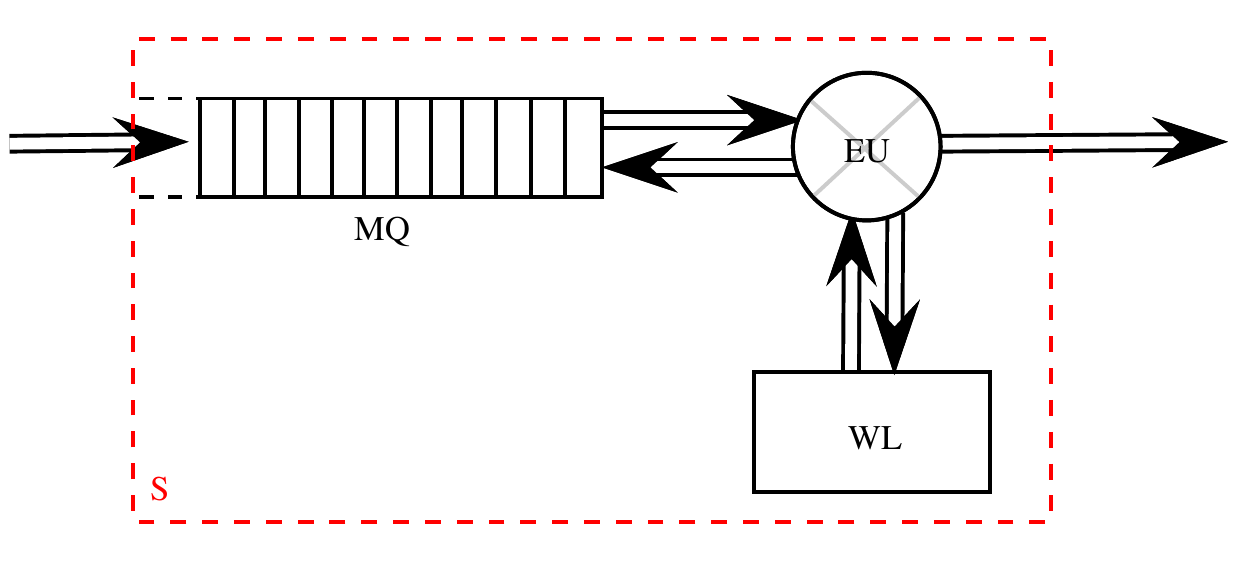}
\caption{Single-queue single-elaboration classical representation of a scheduler.}
\end{figure}
In particular, either in its embryonic or complex structure, the operations involved in the corresponding procedure can be abstracted by using the following logic elements:
\begin{itemize}
	\item Main Queue (MQ) : it's the place where the processes are stored since they enter the server and remain until processed. In case, the MQ can be divided in  sub-queues with different characteristics. 
	\item Elaborating Unit (EU) : it's the place where the processes are elaborated, strictly speaking, and it represents the core of the server.
	\item Waiting List (WL) : if a process needs to be stored during elaboration waiting for a certaing contraint to be satisfied, this is the place where to put it. 
	\item Context Switch (CS) : it is the moment in time at which a change in the elaboration takes place  and the currently served process exits from the EU and a new process enters it.
\end{itemize}
In order to guarantee the completeness of the procedure, we will also consider a Dispatcher (DS), whose duties are to maintain the informations of all processes entering the scheduler updated.
For instance, in common calculators, the DS provides the scheduler with data on each process, such as the total time elapsed from launch, the process' state (running or idle) and the dependencies from other tasks.

Exploiting the elementary concepts just showed, the path of a process inside the scheduler can be described step by step.
\begin{figure}[htb]
\centering
\includegraphics[width=\textwidth]{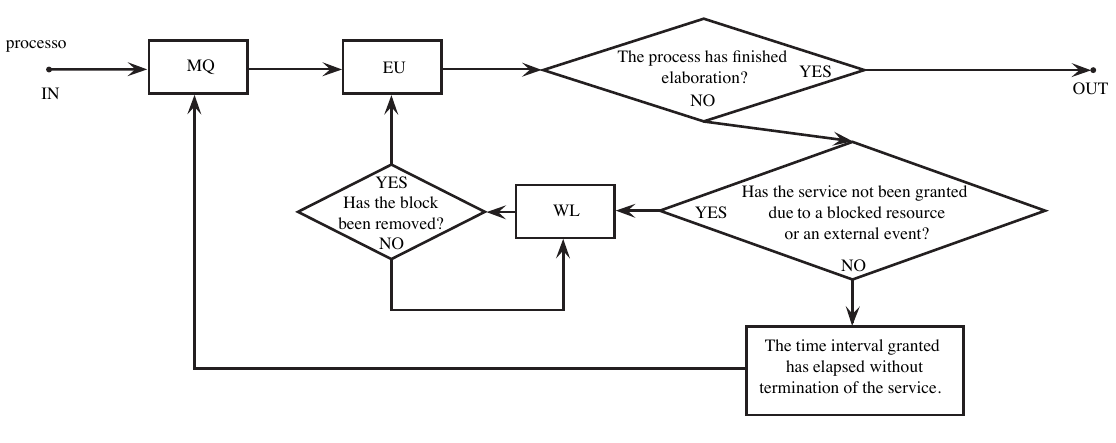}  
\caption{Simplified flux diagram of a Server.}
\end{figure}
Once a process has entered the scheduler device, it is instantly added to the MQ, where it is supposed to remain until served.
If chosen by the scheduling policy, it can leave the MQ, at the next CS, to enter the EU.
As soon as a process is inside the EU, it starts receiving the required service, exiting only in four ways.
It can either reach the end of the time interval granted without finishing the needed service and then get back into the MQ, or if it has finished the elaboration, it exits the scheduler.\\
The other two cases are connected to situation where the job has to be stopped to satisfy a constraint.
These blocks will be referred to as ``interrupts'' and they can be either requested or sustained.
In the former case, the process is added to the WL waiting for a ``resource'' to arrive; as soon as the non operative condition is unblocked, the process is declared ready to be served and eventually elaborated at the next opportunity.
In the latter, the process that was meant to enter the EU is not served, while a process from the WL is elaborated, instead.
This transition can happen as soon as the waiting process is ready (\emph{preemptive strategy}) or as soon as the process currently in the EU exits (\emph{non-preemptive strategy}).\\
For the sake of simplicity, we will assume that each process can enter the scheduler only once at a time, asking for one service only.
More general frameworks can be derived from this scheme by simple additions.

\section{Abstraction: dynamical modelling approach}
In the following we will introduce a paradigm which provides the main elements to derive a parametric model of the relationship between the scheduling strategy and the performances, regardless of the practical implementation.
In particular, we highlight that the system's parameters represent the connection between the scheduler behaviour and the possibly varying computational context.
\begin{figure}[htb]
\centering
\includegraphics[width=0.3\textwidth]{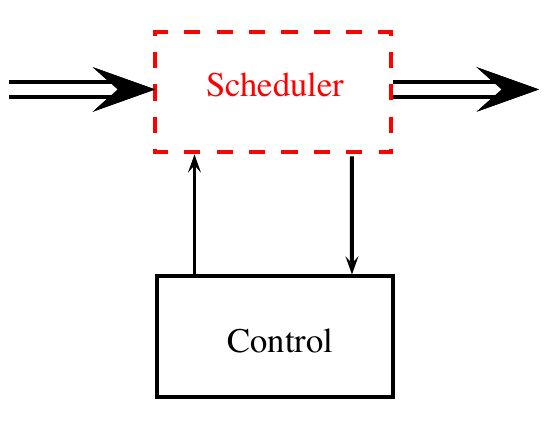}
\caption{Scheduler with the added controller.}
\end{figure}
\subsection{Mathematical formulation}
By means of the simplified scheduling algorithm discussed in the previous section, we will now present the mathematical constructs needed to obtain a control process paradigm.

First of all, the scheduling is a discrete-time process regulated by step of nominal length of $\delta>0$. 
Moreover, let us define the following quantities:
the set of time regulated by the internal clock of the scheduler
\begin{equation}
\mathcal{S}=\{ s \in \mathbb{R} : s=i\delta, i \in \mathbb{N} \}
\end{equation}
and the ordered set of instants at which a CS takes place
\begin{equation}
\mathcal{K}=\{ s_k\}_{k\in\mathbb{N}} \subset \mathcal{S}\quad ,
\end{equation}
which has the property that $s_i<s_j$ ($s_i>s_j$) if and only if $i<j$ ($i>j$).
Then, the length of the $k$-th context switch is 
\begin{equation}
d_k \triangleq s_{k+1}-s_k, \quad s_k,s_{k+1} \in\mathcal{K}\quad .
\end{equation}
Now, it's useful to define the set of all possible processes passing through the scheduler. This set will be identified by $\mathcal{P}$. Let us define also a support function $\sigma : (\mathcal{P},\mathcal{K}) \rightarrow \{0,1\}$ that will be used to perform a test as follows:
\begin{equation}
\sigma(p,s_k)\triangleq \left\{
      \begin{array}{l}
		    1 ~,~ \text{if $p$ is inside the scheduler during $(s_k,s_{k+1})$} \\
		    0 ~,~ \text{otherwise.}
	    \end{array}
      \right.
\end{equation}
Therefore, the set of processes ``inside'' the scheduler at time $d_k$, that is after the $k$-th CS, is:
\begin{equation}
P(k) \triangleq \{p\in\mathcal{P} ~:~ \sigma(p,s_k)=1 \}
\end{equation}
whereas the whole time spent by a process $p\in\mathcal{P}$ in the scheduler is identified by
\begin{equation}
\sum_{k\in\mathbb{N}}\sigma(p,s_k)d_k\quad .
\end{equation}
Any scheduling policy will have a law which chooses the process to be elaborated at the $k$-th CS: we will identify this process by $p^*(k)\in P(k)$.

As far as the treatment has been brought on till now, we have said nothing about how the processes are characterized. There's a great amount of different informations that can be used to represent a process in the scheduler and the followings are only possible alternatives.\\
The most obvious of all is the time spent by a process ``inside'' the scheduler at the time of the $k$-th CS, i.e. $t_i(k)=t^u_i(k)+t^q_i(k)+t^w_i(k)$, where $t^u_i(k)$,  $t^q_i(k)$  and  $t^w_i(k)$ correspond to the time spent in the EU, MQ and WL by the $i$-th process till the $k$-th CS, respectively.
Another classical aspect of a process is its priority, that we will simply identify by $\theta_i(k)$, since it's different for each process and it can even change during time evolution.
Also, the number of times  $\nu_i(k)$  that the service has been blocked for the $i$-th process, waiting for a certain interrupt, until the $k$-th CS may turn out useful.\\
As a matter of fact, the generic process $p_i$ is represented by the above characteristics as a point in a multidimensional space $\mathcal{V}(k)\subset \mathbb{R}^5$, defined by the relation:
\begin{equation}
(k,p_i)\rightarrow V_i(k) \triangleq \left[
\begin{array}{ccccc}
		    t_i^{u}(k) &
		    t_i^{q}(k) &
		    t_i^{w}(k) &
		    \theta_i(k) &
		    \nu_i(k)
	    \end{array}
      \right]^T \in \mathcal{V}
\end{equation} 
We will call $\mathcal{V}$ the characteristic space of processes.

For various operative and computational motivation, it is usually preferred to work with a suitable transformation of $\mathcal{V}$. Thereby, let us define the normalized space $\mathcal{W}\subset\mathbb{R}^m$, $m\le 5$, and the map $f_W ~:~(\mathcal{K},\mathcal{V})\rightarrow \mathcal{W}$, such that:
\begin{equation}
W_i(k) \triangleq f_W(k,V_i(k))
\end{equation}
In order to generalize even more the treatment, we introduce also a partitioning of $\mathcal{W}$
\begin{equation}
\mathcal{C}(k) \triangleq \{C_1(k),~\ldots,~C_h(k) \}
\end{equation}
for which holds
\begin{equation}
\mathcal{W}=\bigcup^h_{i=1} C_i(k)\quad ,
\end{equation}
where
\begin{equation}
C_i(k) \cap C_j(k) =\emptyset ~,\qquad i\ne j \quad .
\end{equation}
Thus, the state of the scheduler at the $k$-th CS can be conveniently described by interpreting $\mathcal{W}$ through the partitioning $\mathcal{C}(k)$, through a proper function $f_X$:
\begin{equation}
x(k) \triangleq f_X(k,\mathcal{C}(k)) \in \mathbb{R}^n
\end{equation} 

It is reasonable, even if not strictly, that the state of the scheduler can be influenced by other elements such as the process chosen by the scheduling policy $p^*(k)$ and the temporal length till the next CS, addressable under the general concept of the input
\begin{equation}
u(k)\triangleq f_U(k,p^*(k),d_k) \in \mathbb{R}^m\quad .
\end{equation}
Then, the update of the state of the scheduler can be formally described through the law
\begin{equation}    \label{eq:control sys}
x(k+1)=g(k,x(k),u(k))\quad .
\end{equation} 
It's important to notice that if the map $f_U$ is bi-univocal, it's possible to decide the scheduling policy in terms of a control problem with regards to the dynamical system \eqref{eq:control sys}.

In the following, we will define the scheduling routine with the purpose of minimizing the mean time that each process spends  waiting inside the MQ, by the design of a proper control law. However, it is possible to perform a similar procedure with regards to others performance parameters. 
\subsection{Functions in practice}
The choice of the map $f_W$, of the partitioning $\mathcal{C}$, as well as of the function $f_X$, decides the form of the state $x$ of the scheduler. In general, there are not optimal choices as such, since the considered state is only an indicator of the dynamic behaviour of the overall scheduler's system, but they depend on the specific problem addressed. In particular, it would be desirable that the obtained state was computationally efficient, but also representative enough of the conditions of the scheduler in that moment, that is able to seize the modifications happening step by step. 
\subsubsection{Choice of $f_W$}
A typical situation in which it is necessary to use a transformation $f_W$ comes true when the time scales are not well fitted to the numerical representation at hand. This happens, for instance, when a subset of the processes can have a very long elaboration (as it is the case of server processes inside a calculator, that need many months of elaboration), whereas others have really short time interval (as the tiny routines performing simple utility functions ).\\
In such a  situation, it would be needed to have a time scale with high precision and, thus, a high computational burden as well. The necessity to employ a map to ``re-normalize'' the space, emerges also from the fact that corresponding state would not be distributed in an uniform way on $\mathcal{V}$. 
\begin{exmp}
To the aim of adapting the situation to a fixed time scale, it's possible to consider only the recent history of the scheduler. In order to introduce the truncation operator, $\left\langle t\right\rangle_{\pi}$, which pulls out the interval concerning the last $\pi$ CS carried out at time $t$, it is possible to define
\begin{align}
	W_i(k) =
	\left[
	\begin{array}{ccccc}
	  \left\langle t_i^{u}(k)\right\rangle_{\pi} &
	  \left\langle t_i^{q}(k)\right\rangle_{\pi} &
	  \left\langle t_i^{w}(k)\right\rangle_{\pi} &
	  \theta_i(k) &
	  \nu_i(k)
	\end{array}
	\right]^T \,.
\end{align} 
$\hfill \square$
\end{exmp}
Generally, the single components $t^{u}$, $t^{q}$ and $t^{w}$ are not distributed on similar scale and so it is deemed reasonable to normalize them on interval of even length.
\begin{exmp}
Let assume that, at the $k$-th CS, there are the following maximum values:
\begin{align}
	& T_k^{u,q,w} \triangleq \max_{p_i\in P(k)}\left\{t_i^{u}(k)\right\}
\end{align}
and let define
\begin{align}
	\bar{t}_i^{u,q,w}(k) \triangleq
	\left\{
	\begin{array}{ll}
		1 & \text{if $T_k^{u,q,w}=0$} \\
		\frac{t_i^{u,q,w}(k)}{T_k^{u,q,w}} & \text{otherwise.}
	\end{array}
	\right.
\end{align}
Therefore, if we choose
\begin{align}
	W_i(k) =
	\left[
	\begin{array}{ccccc}
	  \bar{t}_i^{u}(k) &
	  \bar{t}_i^{q}(k) &
	  \bar{t}_i^{w}(k) &
	  \theta_i(k) &
	  \nu_i(k)
	\end{array}
	\right]^T
\end{align}
the temporal values result normalized in the interval $(0,1)$, even if one of the maximum is null. $\hfill \square$
\end{exmp}
\subsubsection{Choice of $\mathcal{C}$ and $f_X$}
The choice of the partitioning $\mathcal{C}(k)$ and of the function $f_X$ is strongly conditioned by the nature of the processes involved and by the desired dimension of the state $x$.
In general, the ideal would be to keep track of whatever happens to each process inside the scheduler. For obvious computational issues, this cannot be done and thereby it is safer to study the aggregated properties of the set $P(k)$.
With these considerations, the partitioning $\mathcal{C}(k)$ defines the domains to which data should be aggregated and, thus, the choice of the various regions $C_i(k)$ should be designed accordingly, so that each of those is able to gather an homogeneous  set of processes with respect to some criterion. 
This way, $f_X$ has the duty to transform the properties of sets of processes in each $C_i(k)$ into components of the state $x$.
\begin{exmp}
Let assume that the descriptions  $W_i(k)$ of the processes take care only of the normalized times $\bar{t}_i^{u,q,w}(k)$. Therefore, the space $\mathcal{W}(k)\subset \mathbb{R}^3$ is represented by a cube of unit side and with a vertex in the origin.
A simple partitioning that aggregate the processes on the base of their dominant temporal component is given by the eight cubes with side one-half, obtained by putting the new vertexes in the median points of each edge and in the centre of the original cube. $\hfill \square$
\end{exmp}
Within each $C_i(k)$the function $f_X$ can be defined to pick up statistical informations of importance, such as the number of processes, their centre of mass and their dispersion:
\begin{align}
	& \mathrm{Num}\left(C_i(k)\right) =
	  \left\{\text{\# of $W_j(k)$ in $C_i(k)$ at the $k$-th CS}\right\}\in\mathbb{N} \\
	& \mathrm{Cog}\left(C_i(k)\right) = 
	  \sum_{W_j\in C_i}\frac{W_j-G_i}{\mathrm{Num}\left(C_i\right)}\in\mathbb{R}^3 \,,
	  \text{ where $G_i$ is the center of $C_i$} \\
	& \mathrm{Dis}\left(C_i(k)\right) =
	  \sum_{W_j\in C_i}\frac{\left\|W_j-G_i\right\|^2}{\mathrm{Num}\left(C_i\right)}
	  \in\mathbb{R}
\end{align}
where $G_i$ is the center of $C_i$:
\begin{align}
& G_i = \sum_{W_j \in C_i} \frac{W_j}{\mathrm{Num}\left(C_i(k)\right)} \in \mathbb{R}^3
\end{align}
It is obvious that $G_i$ is an aggregated quantity that can be used in turn in the definition of the state.
\begin{exmp}
If we were to put together all these informations to characterize each cube, the final state would have a dimension too high to be elaborated, even if it would be extremely precise. Five values multiplied by eight sets would produce a state with dimension forty! It is thus reasonable to take a few compromises, as the followings:
\begin{itemize}
  \item a single set $C_1(k)$ (that is no partitioning) and exploit of all three precedent values :
    \begin{align}
	    x(k) =
	    \left[
	    \begin{array}{c}
	      \mathrm{Num}\left(C_1(k)\right) \\
	      \mathrm{Cog}\left(C_1(k)\right) \\
	      \mathrm{Dis}\left(C_1(k)\right)
	    \end{array}
	    \right] \in\mathbb{R}^5
    \end{align}
  \item eight subsets $C_1(k),\ldots,C_8(k)$ and their number of processes:
    \begin{align}
	    x(k) =
	    \left[
	    \begin{array}{c}
	      \mathrm{Num}\left(C_1(k)\right) \\
	      \vdots \\
	      \mathrm{Num}\left(C_8(k)\right)
	    \end{array}
	    \right] \in\mathbb{R}^8
    \end{align}
\end{itemize}
$\hfill \square$
\end{exmp}
In the large, the informations needed to characterize the descriptors $W_i(k)$ and, thus, the state $x(k)$ are taken in account by the \emph{dispatcher}.\\
Utilizing aggregated quantity suggests the possibility to employ some \emph{auxiliary descriptors}, that is quantities that are able to add set informations to a certain class. As an example, 
\begin{align}
  \label{eq:medie numeriche 1}
	\mathrm{NU}\left(C_i(k)\right) & =
	  \left\{\text{\# of processes of $C_i(k)$ that are in the EU at the $k$-th CS}\right\}\in\mathbb{N} \\
	\mathrm{NW}\left(C_i(k)\right) & =
	  \left\{\text{\# of processes of $C_i(k)$ that are in the WL at the $k$-th CS}\right\}\in\mathbb{N} \\
  \label{eq:medie numeriche 3}
	\mathrm{NQ}\left(C_i(k)\right) & =
	  \left\{\text{\# of processes of $C_i(k)$ that are in the MQ at the $k$-th CS}\right\}\in\mathbb{N}
\end{align}
contribute to characterize the single class and represent complementary informations to the $W_i(k)$ descriptors.
Anyhow, it's evident that it's possible to generalize the concept of descriptors so to derive such quantities simply by the set functions. Nevertheless, this approach appears formally more complicated and his explication more difficult to comprehend due to the diverse data used together. 
\section{Determining the scheduler's dynamics: the choice of $g$}
Once the nature of the state $x$ and the manipulable input $u$ have been decided, the dynamical function $g$ should be designed to represent the modifications occurring to the first due to the influence of the second.
It is worth observing that such a law depends on the implementation of the scheduler device.
However, generally a detailed modelling always results in a very complex system, whose practical usage turns out difficult.
Furthermore, since the influence of $u$ also depends on the properties and on the flow of the processes, the corresponding system is time-varying in nature.
Hence, a precise derivation of the model from the features of the real device is not suitable to design a performing controller.
Indeed, the dynamical function $g$ has a crucial role in the design of the corresponding control law.
In order to overcome this issue, we may rely on standard identification methods.

If the scheduler is a manipulable device, it is possible to proceed to a preliminary phase of identification.
Given a set of processes to elaborate and their order in the queue, it is established one or more scheduling policy, together with a map $f_U$ which define the related conversion in terms of inputs $u$.
The data so obtained are employed to identify a model, which itself define the $g$.
Here, parametric techniques are preferred, since the system dimensions are already fixed by the choices of $x$ and $u$.
However, different model structures require different identification techniques.
For instance, in the linear system framework, the Recursive Least Square method provides an efficient algorithm to derive the parameters minimizing the square error between the real system and the linear approximating used \cite{kai,mosca,Khalil}.
In general, it is desirable to maintain the dimension of the state $x$ low, in order to deal with a $g$ function and a controller as simple as possible.
Due to this goal it may turn out useful to describe the scheduler's behaviour by various models, each one appropriate to represent a specific functioning phase.
In this case, $g$ is represented by a set of models and correspondently the scheduling policy will have the form of a switching control \cite{libe}.

In case the scheduler could not be manipulated in a strict identification phase, it is possible to proceed adaptively.
Given a structure for the model, simple but flexible enough, it is introduced an on-line mechanism to tune the parameters of $g$ in order to optimize the description of the scheduler just while it is working.
Hence, the corresponding control law turns out adaptive.
In the linear framework, the techniques based on Kalman filtering theory are optimal choices \cite{kai,isi}.

It is important to underline that the identification phase always provides us with a precise information on the model's quality, i.e. on how much narrow the real system's behaviour and the function $g$ are.
In general, there exists a variety of quality indexes, though in standard applications the most used is the mean square error between the outputs of the real and system and its model.
In our situation such a quantity, computed over a time window of $\tau$ CSs, assumes the form
\begin{align}
  e_{\tau}(k) = \sum_{i=1}^{\tau}\left| x(k-i+1) - g\left(k-i,x(k-i),u(k-i)\right)  \right|^2
\end{align}
The quantity $e_{\tau}(k)$ can be fruitfully used both during the design of the model, when the dimensions of the system are chosen, and during its functioning, when a proper threshold could be set in order to provide a fault detection mechanism as well.
However, the analysis of the adherence of the model to the real behaviour and the realization of procedures designed to handle its lost are beyond the scope of this paper.
% stupid informatic's people need explanation
\section{Policy by control strategies}
From a control theory point of view, the majority of standard scheduling algorithms are classified as feedforward techniques, since they realize strategies insensible to the contingent context to handle.
As a matter of fact, they can be regarded as open loop control laws.
Consequently, the corresponding algorithms turn out rather simple to implement, since they rely only onto the present information and, therefore, they have a reduced number of conditional constructs to perform.
However, the corresponding policies can be ``optimized'' in advance in order to ensure the best results only in a single scenario, such as the worst or the most frequent one.
Hence, when the picture is naturally fast changing, these kind of strategies quickly veer off from optimality.

For this reason, in various fields and in particular in that of telecommunications, perfected versions of the standard algorithms have been developed with the aim to introduce simple adaptive mechanisms.
The common idea among all these advanced techniques is based on a simple observation: if you were a priori to know the entire elaboration history of each process, then it would be easy to arrange them in the most convenient way towards the scheduler's performances.
Since this kind of knowledge is not available in advance, in practice it is much more easy to implement elementary techniques based on signal analysis theory in order to update at each CS the processes' classification, assigning the processes themselves a priority index.

Even  though  these strategies are able to introduce adaptive elements, the underlying algorithm are still thought and optimized for precise situations.
This means that the performances of the scheduler will show an improvement only if the variations from the reference context are small.
Moreover, the reactivity of the scheduler towards changes hinges directly on the quality of its forecasts, which in turn tends to increase the computational cost connected with both the dispatcher and the scheduler.\\
The control approach represents the complete evolution of said methodologies, since it lets choose the policy on the basis of a global estimation related to the behaviour of the scheduler, without narrowing the problem of optimization to a few predetermined regimes only.

In the following we will introduce some general ideas to design the next choice in terms of standard control strategies, based on the dynamical model of the scheduler's activity derived by the implementation-free approach of the previous sections.
To this aim we will implicitly assume that the variability of the context is fast changing and then we will focus on adaptive control laws.
\subsection{Self tuning controller}\label{stc}
Hereafter, the main points of this scenario are presented.
A parametric model of the scheduler is first derived along with an admissible region where the parameters are allowed to vary in time.
Then, a suitable controller is tuned by those parameters as they are identified with a proper procedure.

In the following we assume that the scheduling goal can be regarded as a fixed point to be reached in the state space in the least time.
Therefore, the control problem formulation is that of a precise equilibrium point to be globally asymptotically stabilized.
Without loss of generality, we assume that such a point is the origin of the state space.

With regards to the system's parameters identification, we first have to chose a model structure.
For the sake of the simplicity, we use a linear model:
\begin{align}
	x_k = Ax_{k-1} +Bu_{k-1} \quad \forall k \quad .
\end{align}
In order to detect the parameters' variations we could recur to high performance algorithm such as the Recursive Least Square method \cite{kai}.
However, to make clearer the explanation, we will illustrate the traditional formulation in terms of linear algebra.
Hence, let us consider a time window covering a length of $m$ CSs and let us define
\begin{itemize}
 \item $X(k,m) = \left\{x_k,\ldots,x_{k-m+1}\right\}\in\mathbb{R}^{s\times m}$ contains the latest $m$ states
 \item $U(k,m) = \left\{u_k,\ldots,u_{k-m+1}\right\}\in\mathbb{R}^{1\times m}$ contains the latest $m$ inputs
  \item $A\in\mathbb{R}^{s\times s}$ and $B\in\mathbb{R}^{s\times 1}$ represent the matrix to be identified.
\end{itemize}
Therefore, the last $m$ system states can be regarded as
\begin{align}
	X(k,m) = AX(k-1,m) +BU(k-1,m) \quad \forall k,m
\end{align}
that is
\begin{align}
	X(k,m) =
	\left[
	\begin{array}{cc}
		A & B
	\end{array}
	\right]
	\left[
	\begin{array}{c}
	  X(k-1,m) \\
	  U(k-1,m)
	\end{array}
	\right]\quad .
\end{align}
Then, introducing the quantities
\begin{align}
  & V_1(k-1,m) = 
	\left[
	\begin{array}{c}
	  X(k-1,m) \\
	  U(k-1,m)
	\end{array}
	\right] \\
	& M(k-1,m) = V_1(k-1,m) \cdot V_1^T(k-1,m) \\
	& V_2(k1,m) = X(k,m)
\end{align}
if $\det M \neq 0$, it follows that
\begin{align}
	\left[
	\begin{array}{cc}
		A & B
	\end{array}
	\right] =
	V_2(k,m) \cdot V_1^T(k-1,m) \cdot M^{-1}(k-1,m)
\end{align}
Observe that the range of validity of the parameters depends on the chosen model structure.
Roughly speaking, if that structure does not fit the scheduler's behaviour well, the parameters may vary along large intervals.
Thus, it is worth observing that such a case may result an issue in the controller's tuning phase.
However, the analysis of such a situation is beyond the scope of this paper.

Once the parametric model is given, a stabilizing controller leading to the desired state in the least time has to be designed.
To this aim predictive control strategies such as LQ regulators turn out well suited \cite{mosca}.
We underline that the corresponding controllers will depend on the model parameters.
Thus, their performances result as better as slower the parameters' variations are.
This highlights the importance of a good choice for the model structure in order to prevent the parameters from varying their values too much.
\subsection{Model reference adaptive control: Switching Control}   
In MRAC contexts, the behaviour of the scheduler is constantly identified and compared with one or more reference models, in order to evaluate the gap between the models' behaviour and that of the plant and decide the regulation law more appropriate for a controller. In particular, in the scenario of the Switching Control, an array of models is created so that each model is able to adapt to a single operative condition of the system. The number of these different models is higher, the more are the operative states the system can be in. Moreover, each model of the array is associated with a corresponding optimal controller \cite{libe}.

Thus, through an identification phase it's possible to decide which model is appropriate to describe the current behaviour of the system. When this choice is made, the corresponding controller is activated, that is closed in feedback loop on the scheduler itself.\\
From a formal point of view, let us consider $r$ couples $(A_i,B_i)\in\mathbb{R}^{s\times s}\times\mathbb{R}^{s\times 1}$, $i=1,\ldots,r$, and define the quantity
\begin{align}
	e_i(k) = \frac{1}{m}\sum_{j=0}^{m-1}\left\|x(k-j) -A_ix(k-j-1) -B_iu(k-j-1)\right\|^2~,
	\quad i=1,\ldots,r
\end{align}
which define the quadratic mean error of each considered model, calculated at the $k$-th CS on a window of length $m$.
Then, the best representation  of the current state of the scheduler is given by 
\begin{align}
	i^* = \arg \min_{i} \left\{e_i(k)\right\}\quad .
\end{align}
Therefore, the corresponding control law is generally chosen inside a family of functions $f_i(x_k)$, preventively defined on the basis of a performance index. 
Assuming again that the aim consists in stabilizing the system on the origin as fast as possible, it's reasonable to exploit LQ controllers. 
In this matter, for what concerns the speed of commutation between models compared to the performances' variations, it applies analogous considerations to those made in the previous section on the parameters' variability. 
\subsection{Advantages of the control approach}
In the previous section, we emphasized that the scheduling problem can be formulated as a control problem under very general hypothesis.
Then, it is important to understand which advantages one can secure through this new kind of approach compared to more traditional techniques.
It's to be expected, similarly, that the control strategy has better performances when the operative context is characterized by a visible variability, that is, when the situation jumps from regime to regime extremely different among them.
These advantages, however, are obtained at the cost of an augmented complexity of the scheduler and dispatcher alike.
In general, the design of a scheduler according to the control strategy allows to make appropriate choices to adequate the informations handled by the dispatcher, as well as the dimension of the complete system, to specific requests in terms of practical complexity.
Nevertheless, it is intuitive that to obtain better performances it would be necessary to accept a sufficiently high complexity.
In practical issues, usually, it's not immediate a priori to establish if the advantages gained this way are such to justify greater computational and production costs.
Therefore a decision is strictly connected with the specific problem under study.
This is especially true in those cases, such as the CPU scheduling, where the scheduling algorithm is computed by the same elaboration unit.
In this context, augmenting the complexity of the strategy means to increment the time needed to take the next decision and, so, diminish the percentile of time of elaboration given to each processes rather than to the scheduling algorithm.

We would like to introduce some considerations of pure technological nature.
At the same time, in fact, it's important to consider that with the reduction of the electronic products' cost, it's is reasonable to propose a different scenario, in which the elaboration unit is physically and conceptually separated from the scheduler.
The latter, together with its dispatcher for the sake of completeness, become then a device whose operation is parallel to that of the elaboration of processes. Therefore, it is possible to expect that with a convenient design, the scheduler logic could be carried out at the same time of the job of the elaboration unit.\\
Such an alternative can be easily implemented, reformulating the problem \eqref{eq:control sys} as:
\begin{align}
  x(k+1) = g\left(k-1,x(k-1),U(k-1)\right)
\end{align}
In this way, as a matter of fact, in the period between the $k$-th CS and the $k+1$-th CS, all the informations needed to choose the next process are all known and, much more important, this computation can be done in parallel to the elaboration of the actual process.
If the minimum time between two CS is less than the time needed to solve the optimization problem, the two scheduling phases cannot weight upon the net time of elaboration.\\
The advantages of a similar situation are absolutely evident, especially considering complex optimization problems that can be solved by devices with production's costs really low, justifying this way the decision to introduce a separate device. In detail, it is worth to underline that this situation is absolutely congenial in light of the present outlook of electronic computers. Modern CPU, in fact, have two or more cores that can be used to implement calculation in parallel. However, traditional scheduling algorithms are non made to run in parallel architecture and this tend to create small inefficiencies in the operations of single core computer. Vice versa, a technique, as the one described above, is naturally implemented to adapt itself to situations of parallel elaboration.
\section{Numerical example: Adaptive Round Robin}
In order to prove that the above paradigm's elements are sound, it is useful to construct an illustrative example.
To this aim we will refer to the Round Robin scheduling policy, since it has been widely studied in the literature and it represents a simple though solid case of study \cite{joseph,handbook}.
Nonetheless, the following example has not the pretension to describe a real scenario, that is, it does not represent any existing device.
However, it contains all the necessary characteristics we have been talking about and it shows how performances can be improved by means of the control approach.

In order to illustrate the main point of our solution, we have set up a virtual scheduler fed with a flow of processes able to produce two different kinds of stall, already know in literature \cite{bib:book08,bib:book07,joseph,handbook}.
On one hand, there is the stall created by CPU intensive processes, which would require a long time slice to work best.
On the other, there's the stall produced by an enormous number of short life processes, which would require a short time slice to perform optimally.
Since in the Round Robin algorithm the only parameter is the time slice's length, the problem is utterly reduced in dimension, providing a simple but illustrative scenario.

In this example, we have decided not to analyse any details concerning operations that the scheduler must perform beforehand, such as the deadlock avoidance algorithm, assuming that they have mechanisms of their own.
However, being a kind of Round Robin algorithm, the resources physical deadlock can be handled efficiently with common techniques \cite{bib:book08,joseph}.
In our case, we have simply decided that each job seizes all the needed resources before being processed, releasing any resource upon leaving the EU. This way, at each CS, the scheduler has simply to control if between the processes inside the WL, there is at least one for which the waiting condition has ended, in order to pick it up from the idle condition and put it back into elaboration. \\
This said, the only control needed to avoid deadlock is the presence of a circular waiting between processes inside the WL. For the sake of simplicity, in our simulative context we have a priori imposed the coupling relation between processes, in order to prevent the possibility of circular waiting.
\subsection{Round Robin}
The Round Robin (RR) policy is one of the simplest scheduling algorithms for tasks in an operating system, which assigns time slices to each job in equal portions and in circular order, handling all processes without priority. Round Robin scheduling is both simple and easy to implement. In order to guarantee starvation free runs, the Round Robin is implemented with an ageing policy, that prevent that CPU hungry processes can monopolize it.

\begin{figure}[htb]
\centering
 \subfigure[\small{Mean Elaboration Time}]
  {\includegraphics[width=0.45\textwidth]{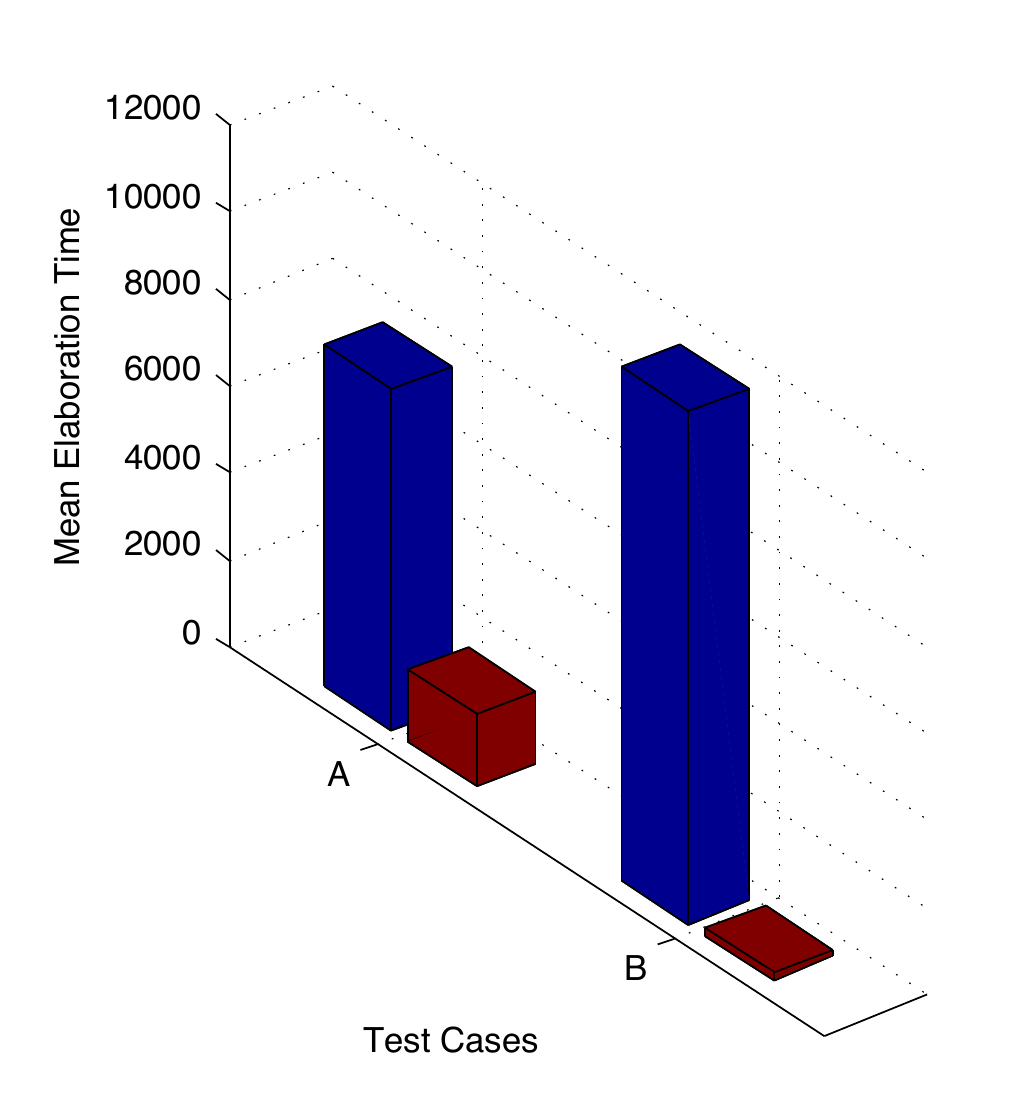}}
 \hspace{0mm}
\subfigure[\small{Mean Number of Interruptions}]
{\includegraphics[width=0.45\textwidth]{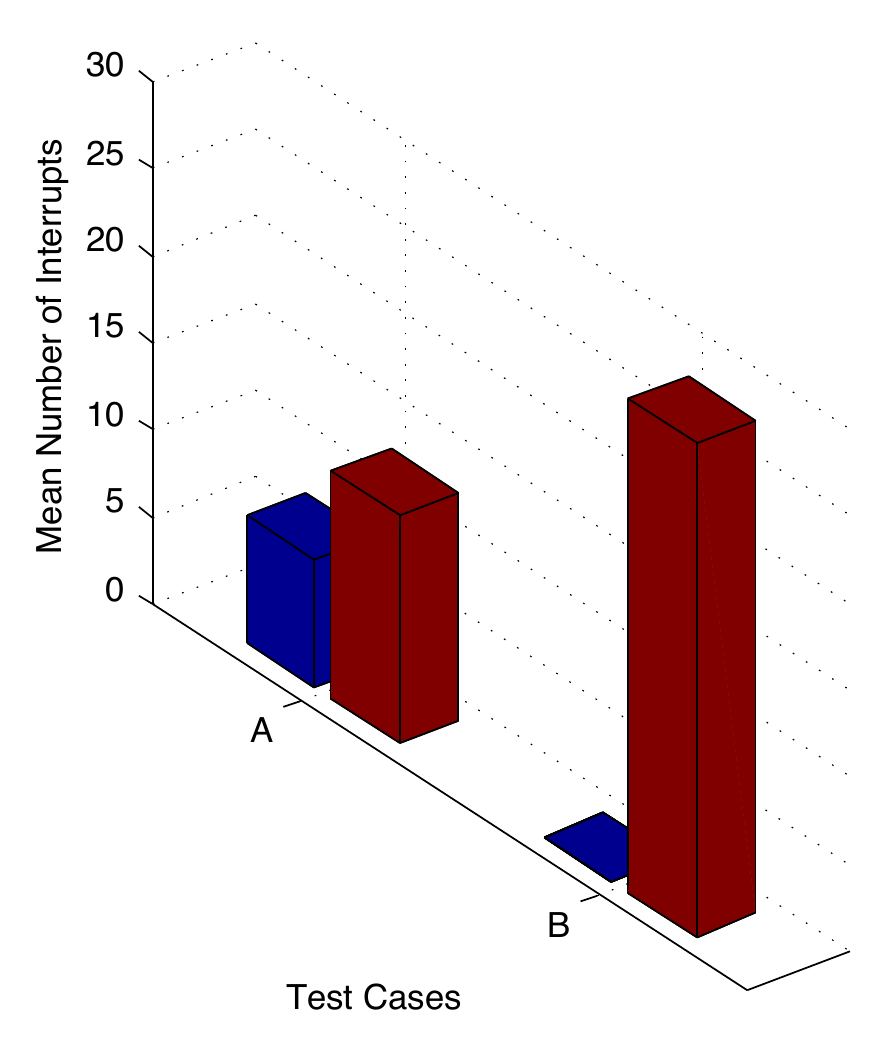}}
 \hspace{0mm}
\subfigure[\small{Mean Interruption Time}]
{\includegraphics[width=0.45\textwidth]{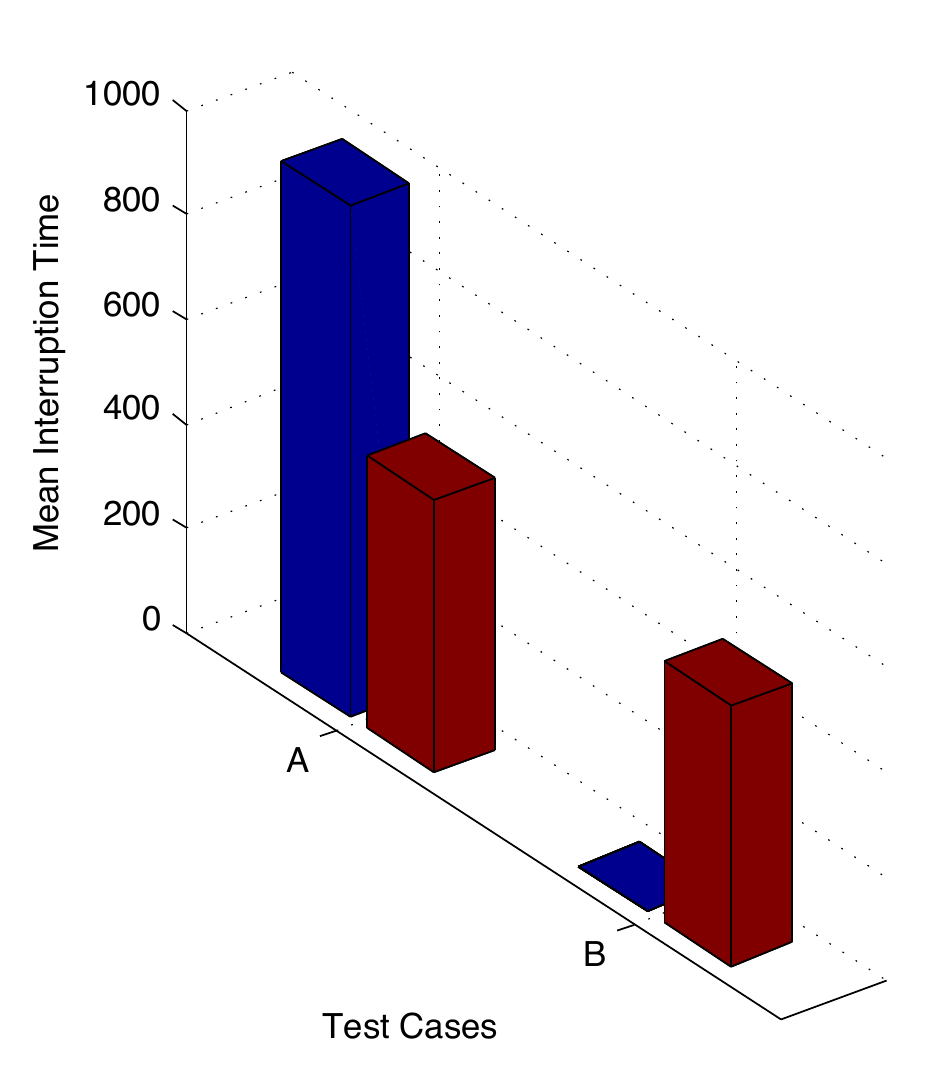}}
\caption{\small{Histograms of the statistical properties of the two situations of stall A and B. Both are characterized roughly by two different task generators.} \label{isto}}
\end{figure}

\begin{figure}[htb]
\centering
 \subfigure[\small{Test Case A}]
  {\includegraphics[width=0.45\textwidth]{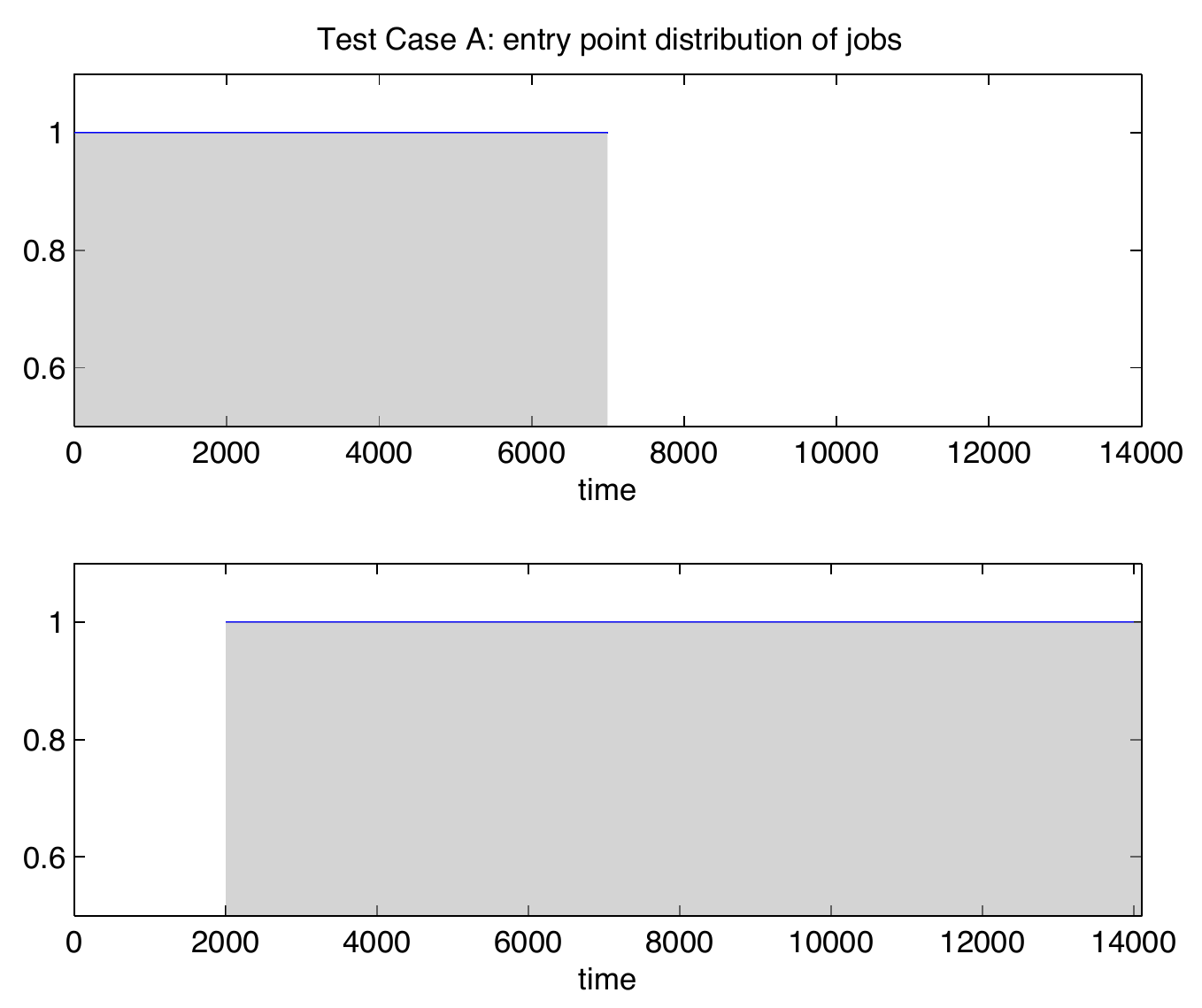}}
 \hspace{0mm}
\subfigure[\small{Test Case B}]
{\includegraphics[width=0.45\textwidth]{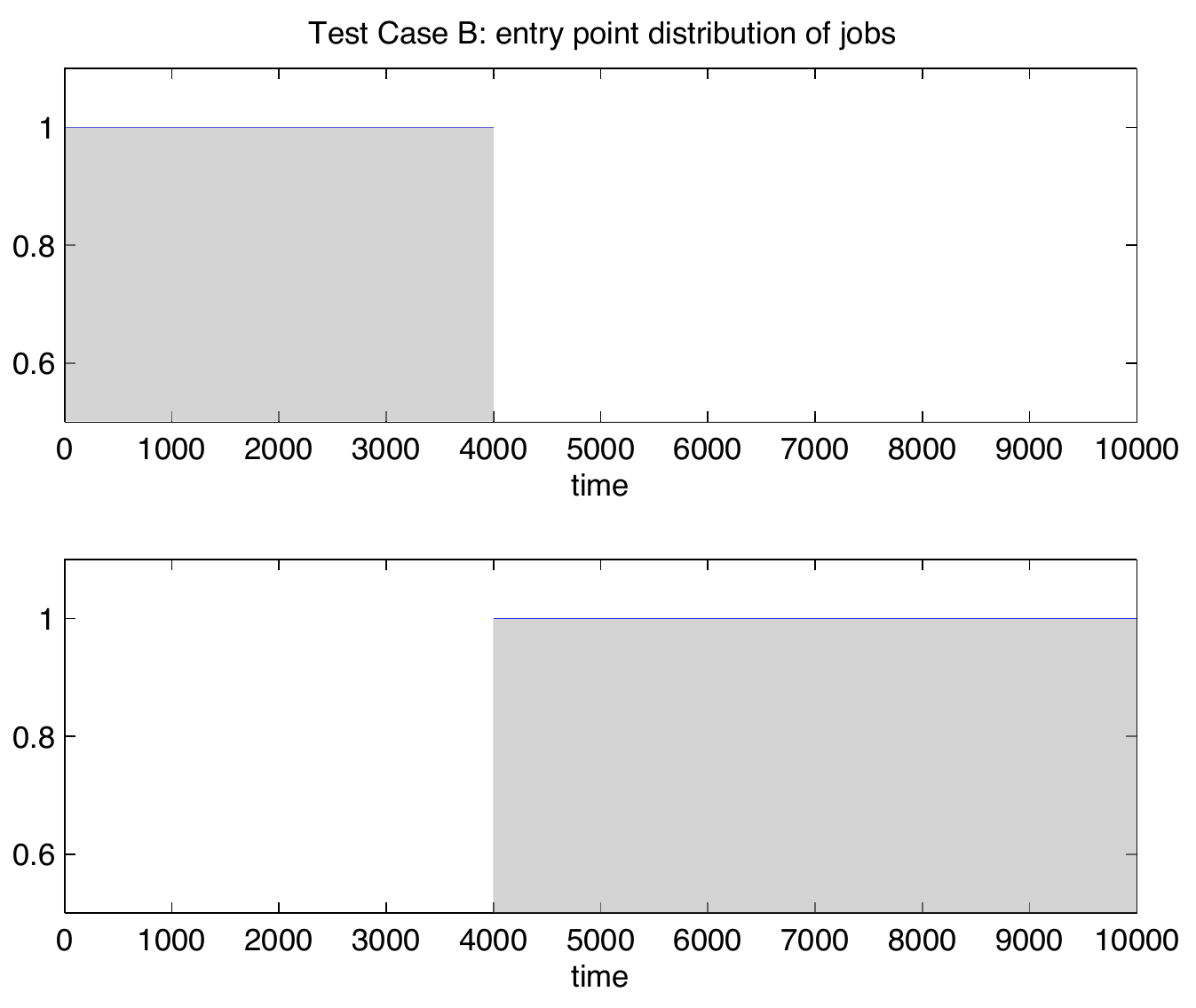}}
\caption{\small{Entry point distributions of jobs in the two test cases A and B, with regards to two different task generators.} \label{entry}}
\end{figure}

As it is seen, the only control variable we have access to is the time slice. This is important because it'll diminish the dimension of the problem and lets lead a more in-depth study.
In order to start the analysis, it's important to set up the scenarios in which  to test the behaviour of the scheduler. With regards to the RR's scheme, a few simple stall's (performances' degeneration) conditions have been detected. In particular, in the example proposed, the following test cases have been considered:
%Distinzione casi di stallo - allungare itemize
\begin{itemize}
\item CPU intensive processes (case A): this is the situation in which the majority of the processes are elaborating-like routines. Those processes are characterized by a long elaboration time which needs more than a single time slice to be served. 
\item Spike processes (case B): these processes can be served with short time slice but they usually come in great numbers.
\end{itemize}
%considerazioni sulle prestazioni
In order to obtain the previous two situations of stall, a flux of processes characterized by the statistical properties, depicted in Figure~\ref{isto} and Figure~\ref{entry}, have been used.

\begin{figure}[htb]
 \centering
 \subfigure[\small{A and B test cases}]
  {\includegraphics[width=0.45\textwidth]{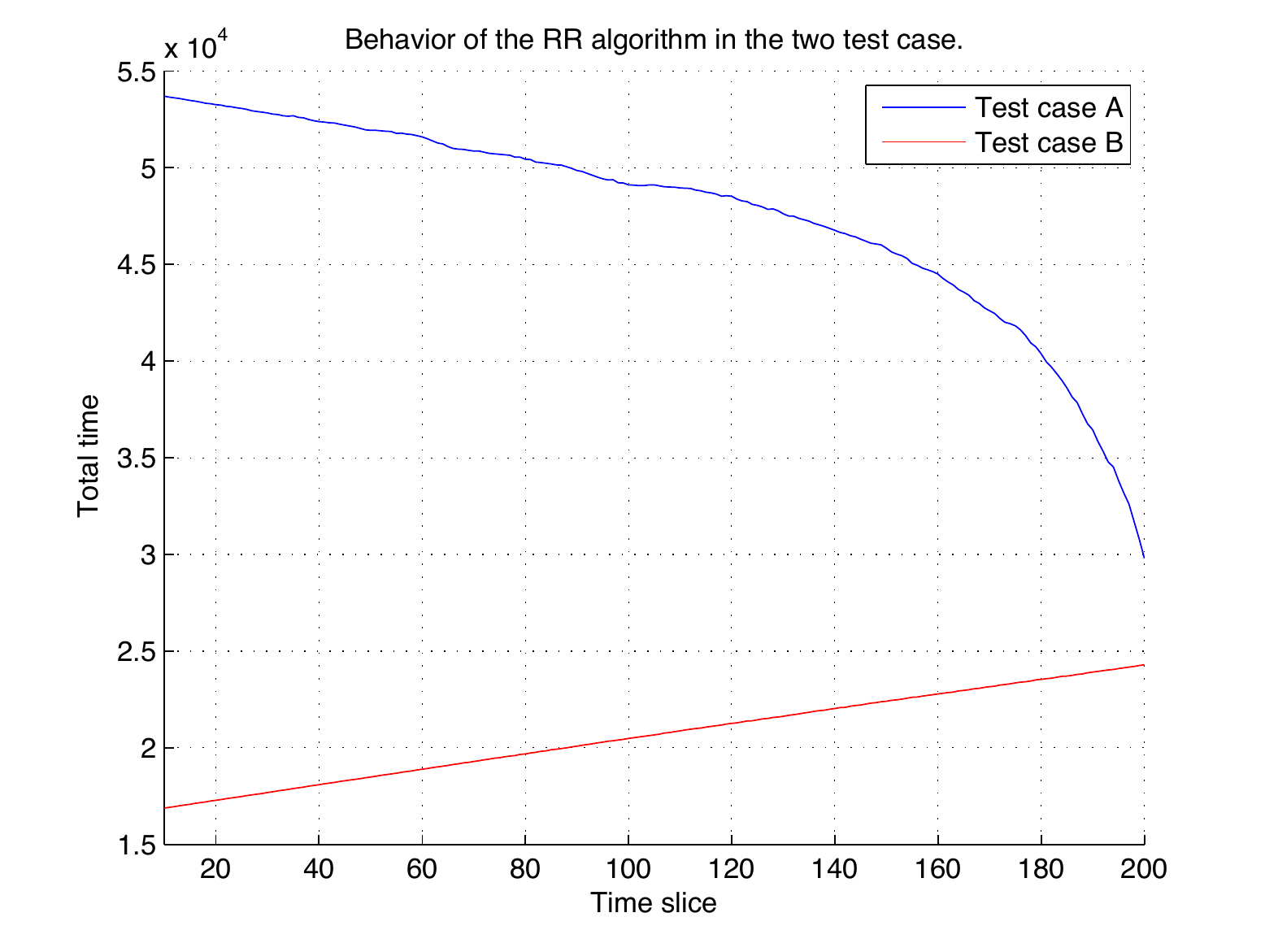}}
 \hspace{0mm}
\subfigure[\small{Normalized test cases}]
{\includegraphics[width=0.45\textwidth]{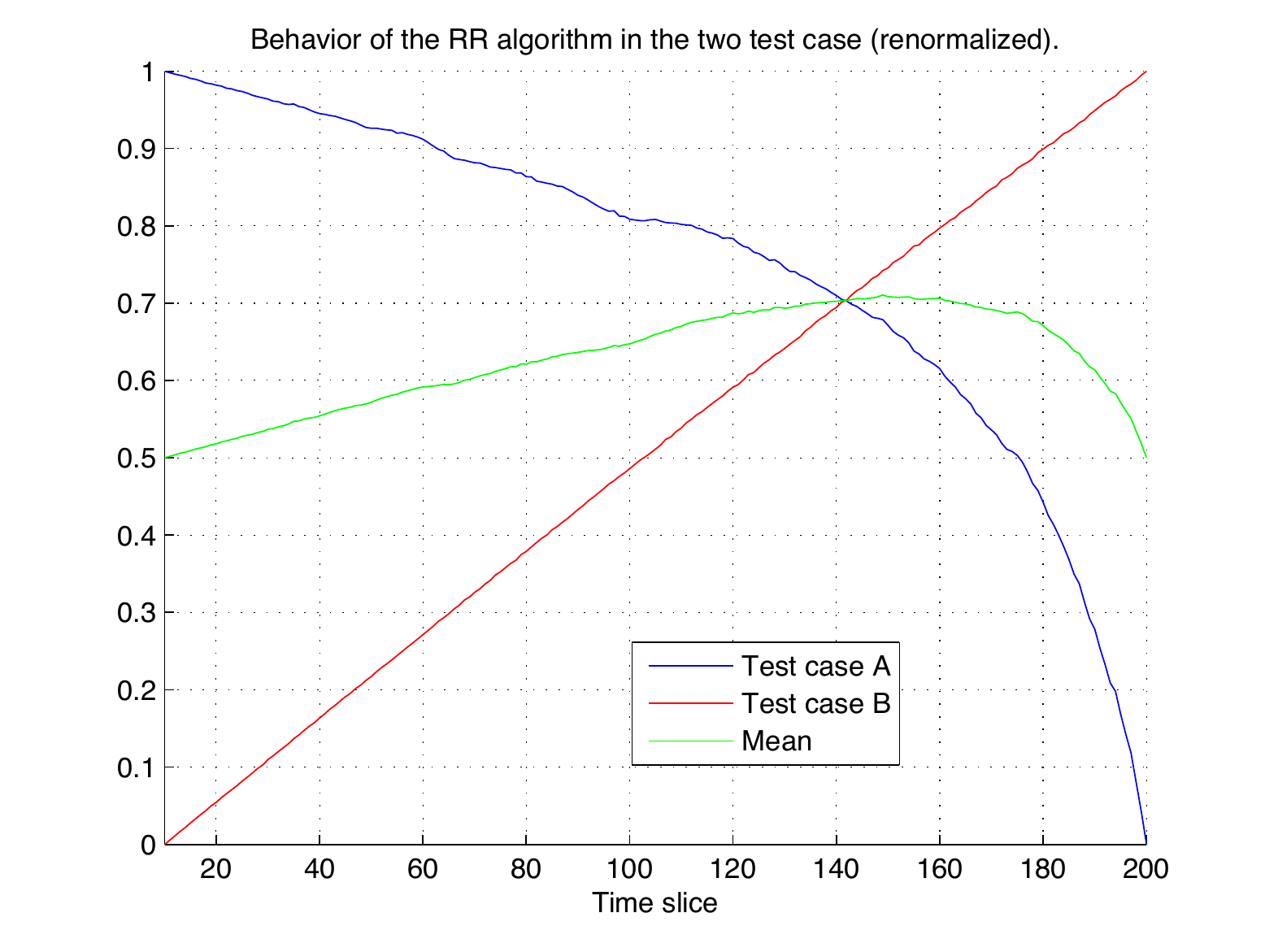}}
\caption{\small{(a): Operative behaviour of the RR algorithm in the two test cases. (b): Normalized behaviour of the RR in the test cases A and B, and the worst case scenario. The normalization is obtained considering that the value 1 is the worst result (where the scheduler has made the worst choice at most steps), whereas 0 is the best result (where the scheduler has made the best choice at most steps).}\label{A+B}}
\end{figure}

%Osservazione diretta sulle prestazioni del RR in presenza di casi di stallo diversi
Looking at Figure~\ref{A+B}, it's observed that the performances of the RR algorithm are maximized for the respective optimal case. The elaboration context of the test case A would benefit by a long time slice, whereas, in the test case B, the best choice is to use a short time slice. 

Observe that in the case A, a long time slice guarantees a time gain equal to $3,4\times10^4$ compared to the short time slice, while in the B case a short time slice guarantees a gain of $0,8\times10^4$, compared to a long time slice. Considering then some alternation even stochastic, of the two limit cases, it's reasonable to obtain that the performances of the scheduler are influenced more by the case that appears more frequently, with respect to the ratio of the two gains.  
In the example provided, the frequency of each case is weighted by the reciprocal of the corresponding gain. In this way, the choice of a long or short time slice gives on average the same result. In Figure~\ref{A+B}, the performances of intermediate time slices are presented alongside the previous single cases' ones, on a normalized scale.
%Contesto peggiore, A e B si alternano in maniera stocastic

%\includegraphics[width=0.8\textwidth]{A+B+mean.pdf}    
% nella caption di questa figura è d'uopo sottolineare il metodo di normalizzazione

\subsection{Dispatcher}
A dynamic dispatch is the process of mapping a message to a specific sequence of code (method) at runtime. This is done to support the cases where the appropriate method cannot be determined at compile-time (i.e. statically). Dynamic dispatch is only used for code invocation and not for other binding processes (such as for global variables) and the name is normally only used to describe a language feature where a runtime decision is required to determine which code to invoke.

Obviously, a dispatcher is always present in any scheduler in order to save and load all the informations needed for elaboration. In our two cases below, a few differences had to be introduced and so the dispatcher has been changed too.

 Note that the choice to calculate in advance some temporal quantities relative to the last $\pi$ CS is motivated by the choices made for the scheduler's realization.
 
 The computational cost of the dispatcher is clearly $\mathcal{O}(n)$, that is linear with the number of active processes once fixed the $k$ parameter (the Appendix contains the simplified code's lines for the modified dispatcher). This result is exactly the same as the one of the current operating system.
 
\subsection{Adaptive Control}

As it has already been pointed out in the Section~\ref{stc}, in order to apply a suitable control policy to the scheduling, the control system has to know an operating model of the dynamic of the scheduler itself. Thus, a suitable model has to be deduced.

First of all, we have to derive a suitable state vector of the scheduler.
To this aim, we can group all the processes in a single class, which we will calculate a series of aggregated quantities for, sufficiently representative of the whole set. To avoid that processes evolving on a too long temporal scale (such as server processes) can introduce unwanted deviations in the set means, we will compute all the necessary values with regard to the latest $\pi$ CS.
%Recalling that the dispatcher is already designed with this problem in mind, we can observe that the quantities  \verb|container.meanTU|, \verb|container.meanTW|, \verb|container.meanTQ| are equivalent to the first three components of $G_i$ in  \eqref{eq:medie temporali d'insieme}, while \verb|container.meanNU|, \verb|container.meanNW|, \verb|container.meanNQ| correspond to the values in \eqref{eq:medie numeriche 1}-\eqref{eq:medie numeriche 3} mediated on a time window of $\pi$ CS.

In the following example, we propose a certain state vector, through some of the representative quantities of the computational context of the scheduler that can be found, for instance, in \eqref{eq:medie numeriche 1}-\eqref{eq:medie numeriche 3}. Due to the issues of normalization and numerical optimization  we have modified the chosen variables through monotone function. Anyhow, we would like to stress that the dispatcher, which we have already spoken of, presents numerous alternatives to the construction of plausible state vectors. 
The chosen quantities are :

\begin{itemize}
\item $T\%_u$ :  percentage of the total time spent in the EU by all processes;
\item $N_q$ : number of processes in the MQ;
\item $\delta{N_q}$ : variation of the number of processes in the MQ.
\end{itemize}
thus, the state vector is  as follow,

\begin{align}
x=\left[ T\%_u\quad N_q\quad \delta{N_q}\right]\quad .
\end{align}
All the mean quantities evaluated just cited have been calculated over a temporal window of length 32 CS.

It is important to note that even if the operations between arrays can be computationally taxing, in our case the dimension $c$ of the state is fixed as much as the width $m$ of the temporal window. With the two parameters $c$ and $m$ chosen, the computational cost is constant independently  from the number of processes introduced, that is it grows as $\mathcal{O}(1)$ considering the notation used in the computer language. 

It is obvious that the choices of this values affect directly the capacity of the scheduler to identify correctly a suitable model. The respective code can be found  in the Appendix, only to highlight this aspect, knowing well that to compute a linear regression there are far more refined algorithms.

\begin{figure}[htb]

 \centering
 \subfigure[\small{Test case A}]
{\includegraphics[width=0.45\textwidth]{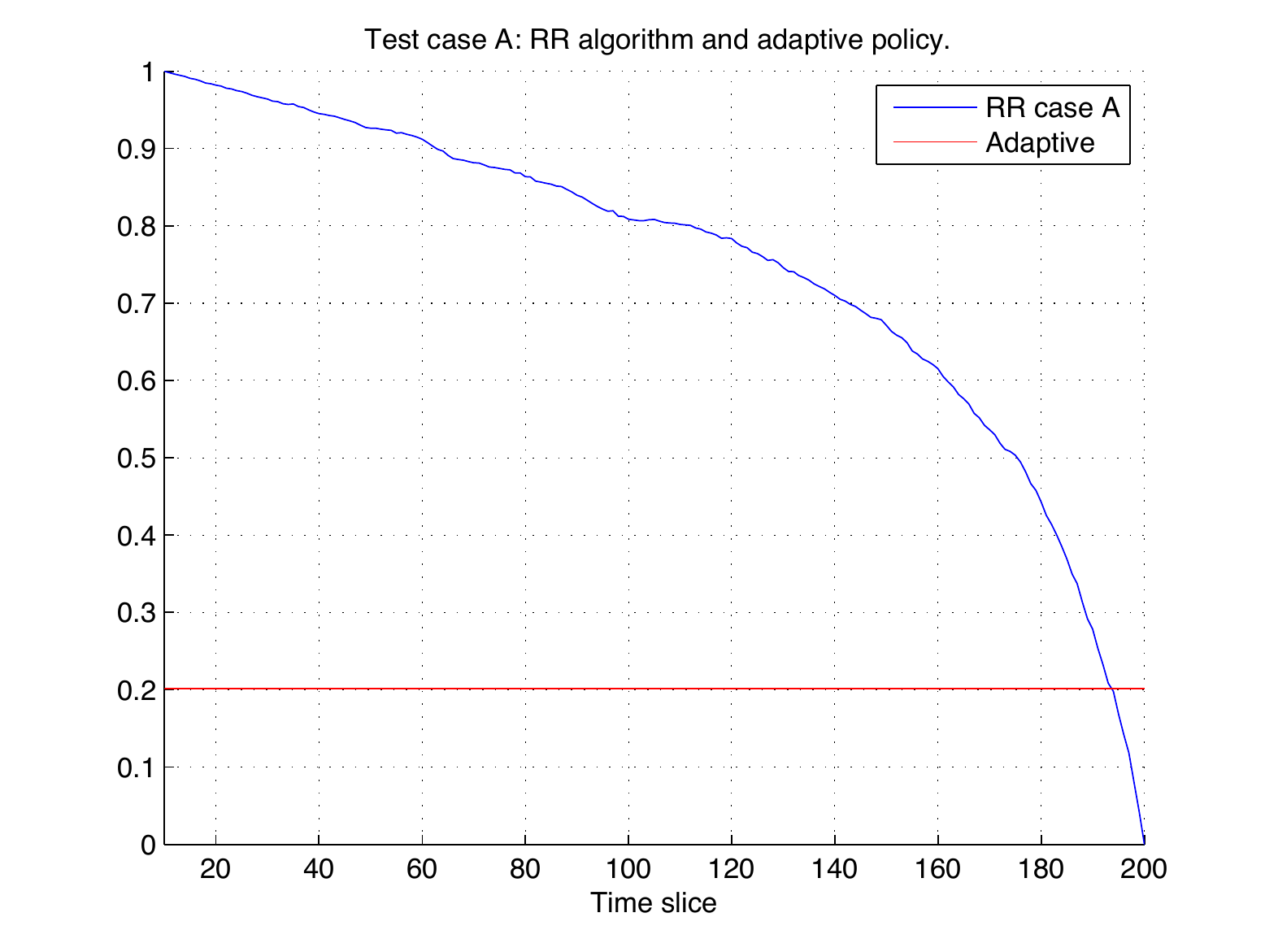}}
\hspace{0mm}
 \subfigure[\small{Test case B}]
 {\includegraphics[width=0.45\textwidth]{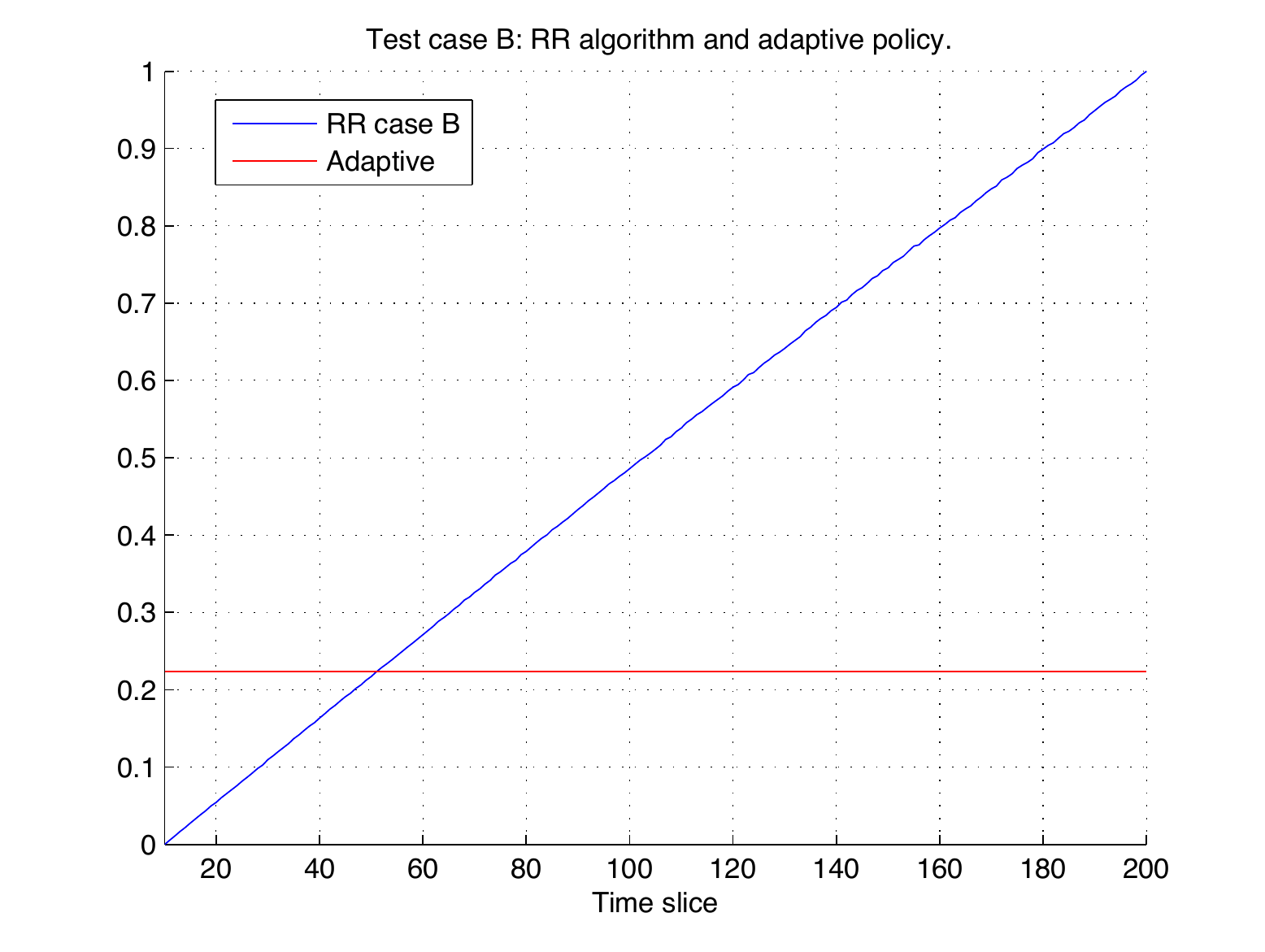}}
\caption{\small{Test case A and B: confront between the results obtained with the RR and the Adaptive algorithm.}\label{AB+a}}
\end{figure}

On the basis of the identified model, it is possible to predict the value of the next state and so the next time slice. To simplify the treatment of our example, we decided to build a state to guarantee that the more the state is close to the origin, the more the scheduler is far from a computational stall situation.

In this case, then, we have to minimize with respect to the choice of $u_k$ the following quantity:
 
 \begin{align}
	\left\|x_{k+1}\right\|^2 =
	x_{k+1}^Tx_{k+1} =
	x_k^TA^TAx_k + 2x_k^TABu_k + B^TBu_k^2
\end{align}

We underline the fact that the nature of the Round Robin itself and the presence of anti deadlock starvation checks guarantee that this minimization is not constrained, as in the common control problems, where many properties have to be established.
Therefore, if  $B^TB\neq 0$ the solution of the problem above is given by:

\begin{align}
	u_k = -\frac{x_k^TAB}{B^TB}
\end{align}

Since $u_k$ is a time or, more appropriately, the length of the next time slice, it is bounded to assume values inside a certain interval $u_k\in(t_{\text{min}},t_{\text{max}})$, for $0<t_{\text{min}}<t_{\text{max}}$.

To the aim of showing results rigorously derived from the theory, much more than a possible arbitrary implementation, the decisions taken for the code (see the Appendix) are true to the formulas shown before (Section~\ref{stc}). It is important to stress that there exist other more efficient implementations of linear regression\footnote{The most famous of these techniques is the RLS (\emph{Recursive Least Square}).}. In the case under exam, the regression's window has been set to 64 CS.

\begin{figure}[htb]
\includegraphics[width=0.8\textwidth]{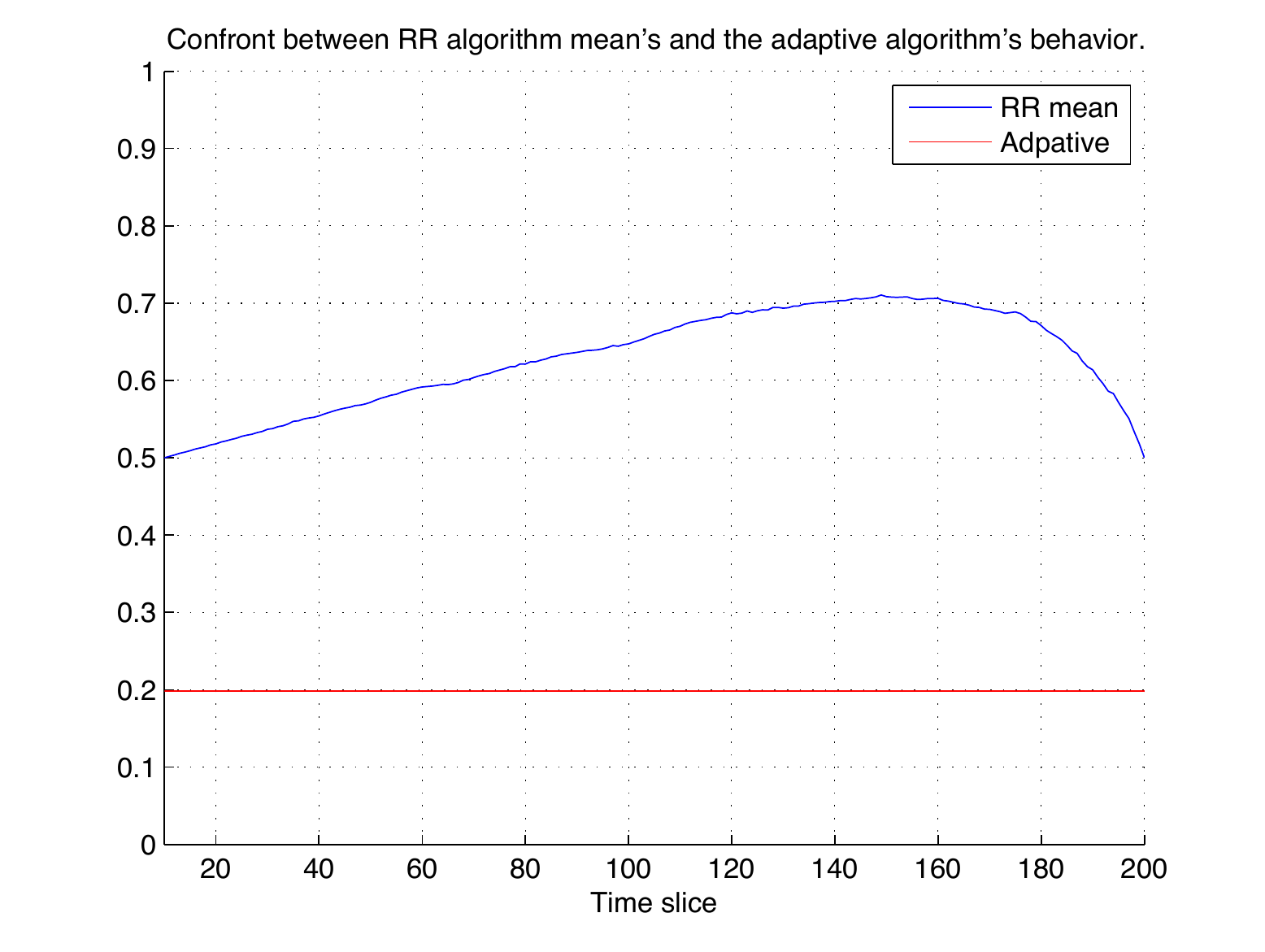}
\caption{\small{Worst case scenario: confront between the results obtained with the RR and the Adaptive algorithm.}\label{M+a}}   
\end{figure}

The following figures show the performances, considering the single stall cases (Figure~\ref{AB+a}) and with respect to the worst case scenario, in which the test cases A and B alternate continuously (Figure \ref{M+a}). 
%%%%%%%%%%%%%%%%%%

\subsection{Switching Control}   
%Switching
%Identificazione classica -> ricavato 4 modelli, buoni per situazioni di comportamento diverse
%Legge di switching
Exploiting the same state vector of the previous approach, the structure of the parametric representation of the scheduler has been set.
Then, through an initial classic identification phase of the input-output type, four models have been constructed, able to describe the four most characteristic phase of operation that are evident in the worst case scenario. The parameter used in the switching law is simply the mean quadratic error committed by the four models, computed over a window of 32 CS.

\begin{figure}[htb]
 \centering
 \subfigure[\small{Test case A}]
{\includegraphics[width=0.45\textwidth]{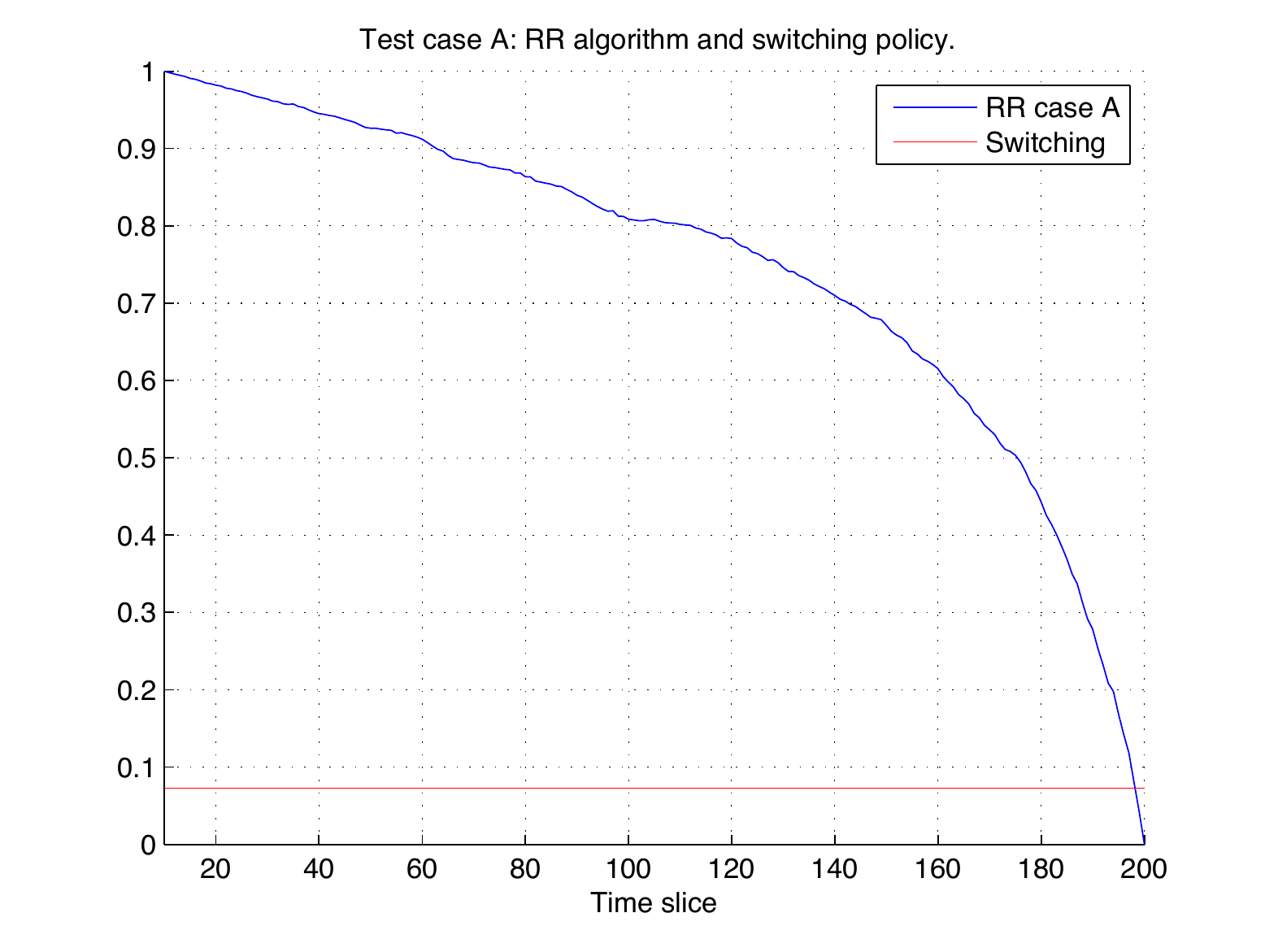}}
\hspace{0mm}
 \subfigure[\small{Test case B}]
 {\includegraphics[width=0.45\textwidth]{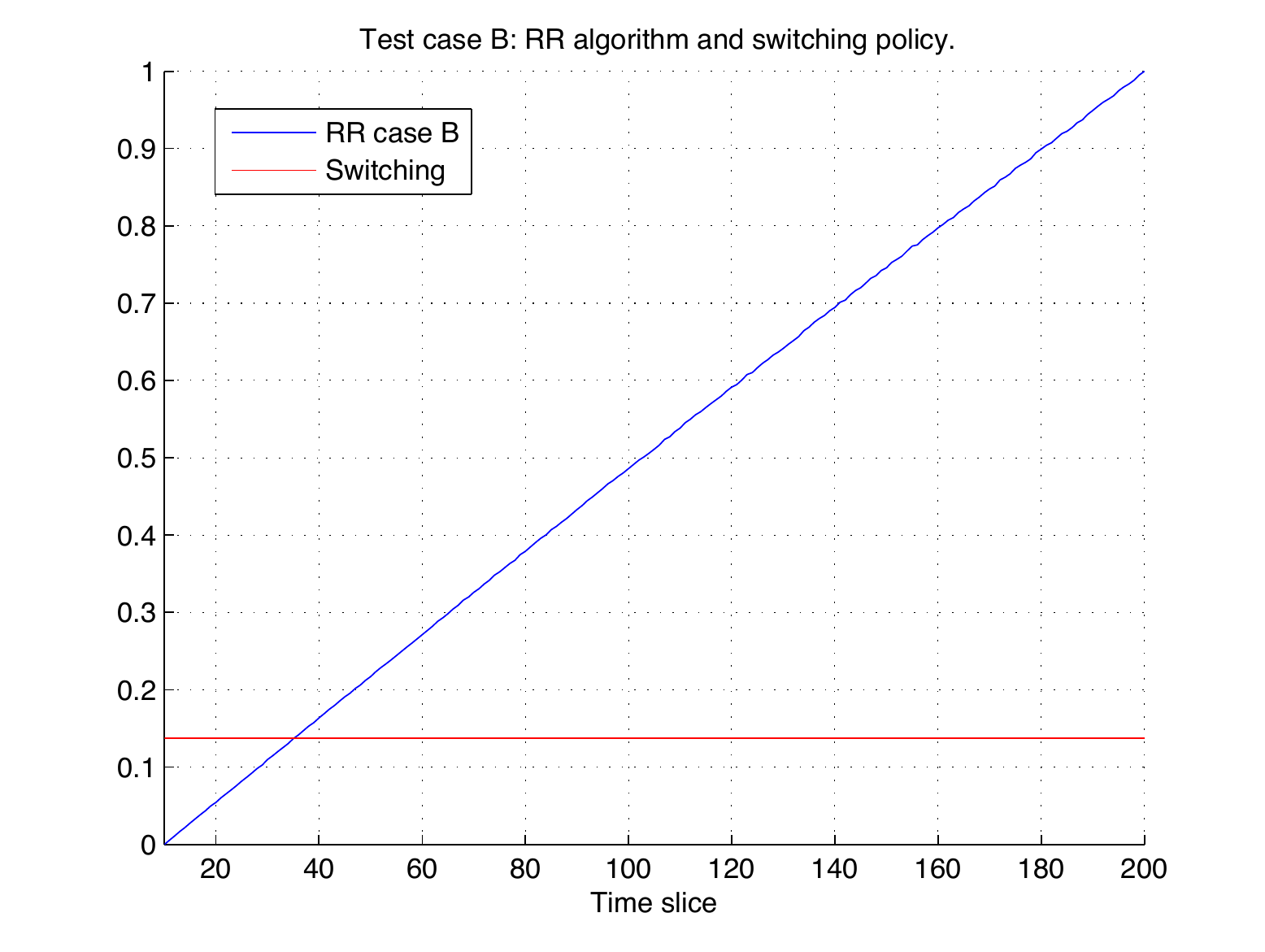}}
\caption{\small{Test case A and B: confront between the results obtained with the RR and the Switching algorithm.}\label{AB+s}}
\end{figure}

\begin{figure}[htb]
\includegraphics[width=0.8\textwidth]{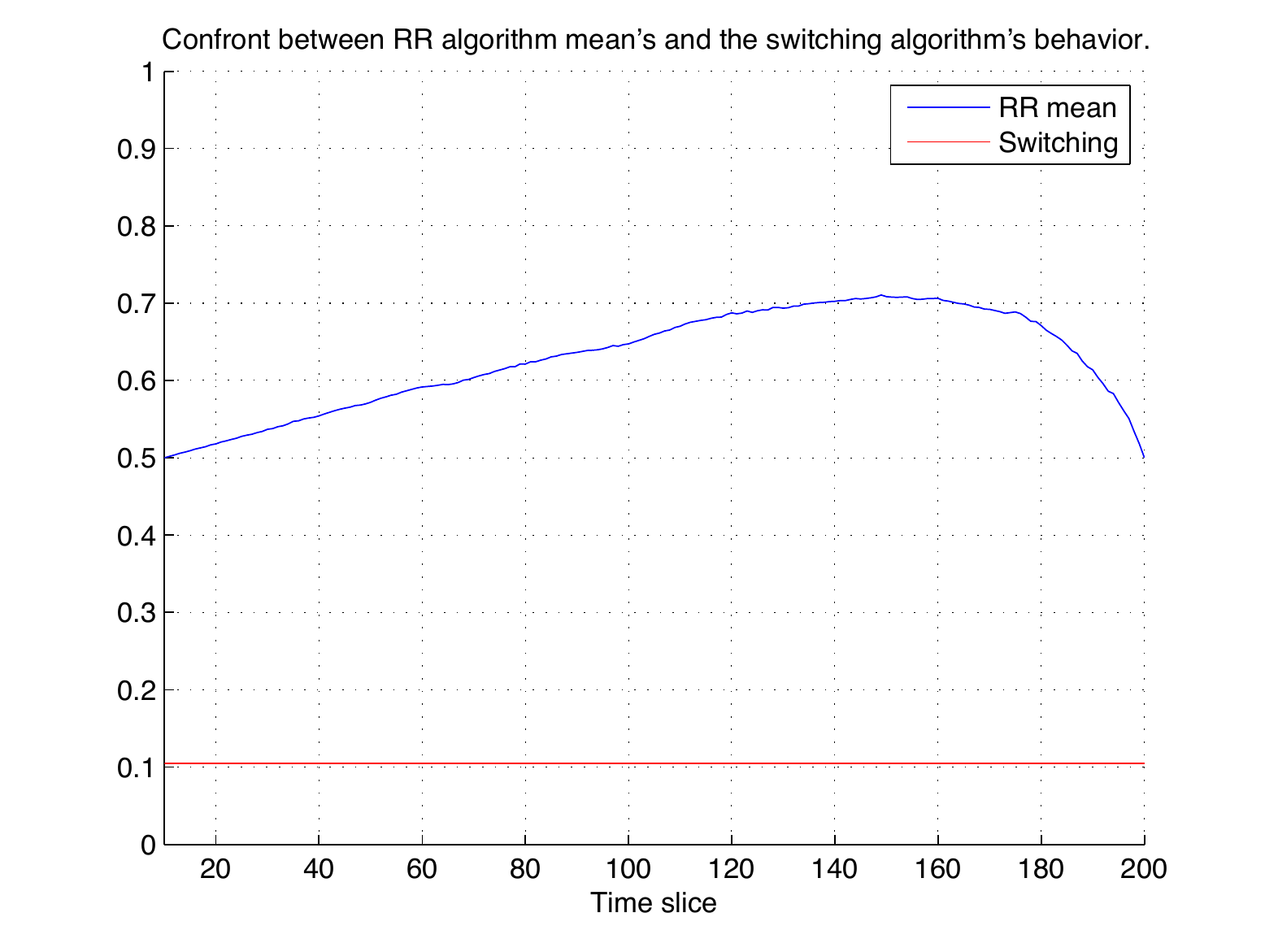}
\caption{\small{Worst case scenario: confront between the results obtained with the RR and the Switching algorithm.}\label{M+s}}  
\end{figure}

In this case too, the figures show the performances with respect to the single stall cases (Figure~\ref{AB+s}), as well as with respect to the worst case scenario (Figure~\ref{M+s}).

\subsection{Remarks}

Before concluding, we would like to point out a few brief technical remarks, that we consider important to the general case more than for the example shown. The two chosen strategies have both good points and bad points. Knowing the advantages or disadvantages of a particular technique is important to be able to minimize the disadvantages.

The adaptive strategy discussed above is computationally taxing, due to the matrix inversion needed to compute the control law. Whatever the function used to calculate the inverse of a matrix or even the product between matrices, an high dimension state is still risky, even if the hardware would consent high performances all the same. On the plus side, though, the Adaptive control is able to shape a larger variety of situations with a small number of informations on the system to control. 

The switching strategy, on the other hand, has the necessity of a deeper understanding of the operative behaviour of the system in order to produce visible results. Once this need is fulfilled, however, one can expect that the switching strategy could work very well at a reduced complexity cost than the adaptive one.

Lastly, we would like to stress that the CS has its own time length. If all the procedures for the control strategy to work properly can be computed in that time, the control strategy does not introduce any delay. We have taken into account this issue while preparing the example, pondering, however, that any real control technique would benefit from a stand-alone implementation rather than to share the EU runtime.

\section{Conclusions}
In this paper we have inquired the problem of deriving an adaptive
policy, when the scheduling problem is settled in a time variant
context.
We have also assumed that the number and the nature of the users or
tasks to be processed can only be described in term of statistics,
though their representation as stationary stochastic processes is not
feasible.
In such a framework we have illustrated a method based on systems
identification theory to abstract the scheduler in order to derive a
proper time varying model.
Then, exploiting adaptive control techniques we have provided a
suitable policy aimed to optimize a desired performance index under
certain constraints.\\
The above technique have been used to design an improved version of
the traditional Round Robin algorithm.
By a numerical example it has been pointed out that in a time variant
framework there isn't an optimal static choice for the time slice.
In turn, allowing the time slice to vary according to a proper adaptive
law, it is possible to improve the scheduling performances by a large
amount.

\newpage
\appendix

\section{Implementation code}

\subsection{Introduction}

The informations here presented are by no means intended to represent the real implementation of any part of a scheduler or of its possible controller. It is well known by the authors that improved and refined routines can be found even in literature to solve efficiently many of the following functions. Nonetheless, in order to show the idea behind the results and to guarantee an higher understanding of the theory, the code is presented in its more simple form.

\subsection{Objects and Functions}
\textbf{\texttt{List} class:}\\
this class handles lists of objects of the types used in the code of the dispatcher and of the scheduler; in particular, here are a few utility functions:
\begin{itemize}
\item \texttt{list.del(i)}: elimination of the $i$-th element from the list 
\item \texttt{list.append(element E)}: addition at the bottom of the list of a new element \texttt{E} 
\item \texttt{list.getArray()}: through this utility function, a list \texttt{list} of numbers or of arrays of dimensions $a\times b$ returns an array formed by the concatenation of the arrays  or of the numbers that constitute its elements
\item \texttt{new list(list H)}: new list type element \texttt{list} can be initialized with copies of other lists\end{itemize}
\textbf{\texttt{Proc} class:} \\
it describes all the informations commonly used by operative systems to handle the elaboration of a process, such as the process' state, the resources allocated as well as any flag triggered by dependencies from or by other processes. We add to this generic class various other informations connected with its past elaboration history:
\begin{itemize}
\item \texttt{proc.U}: total time spent in the CPU
\item \texttt{proc.W}: total time spent in the Waiting List
\item \texttt{proc.Q}: total time spent in the Main Queue
\end{itemize}
\textbf{\texttt{Container} class:} \\
 it is  a class which contains the list of all active processes at a certain time instant and it provides a number of functions suitable to obtain the informations needed to the scheduler's activities. We show a few functions useful to the comprehension of the rest of the code
 \begin{itemize}
  \item \texttt{container.getID()}: it returns a list with all the processes enumerated by ID
  \item \texttt{container.getProc(ID)}: it returns a ``proc'' object correspondent  to the chosen ID 
  \item \texttt{container.getn()}: it returns the total number of processes inside the scheduler
  \item \texttt{container.Ulist}, \texttt{container.Wlist}, \texttt{container.Qlist}: they are lists of the \texttt{list} type used to store the mean time in CPU, WL and MQ in the latest $\pi$ Context Switch ($\pi$  a priori fixed)
  \item \texttt{container.NUlist}, \texttt{container.NWlist}, \texttt{container.NQlist}: they are lists used to store the number of processes in CPU, WL and MQ in the latest $\pi$ CS
  \item \texttt{container.meanTU}, \texttt{container.meanTW}, \texttt{container.meanTQ}: global mean time  in CPU, WL and MQ in the latest $\pi$ CS
  \item \texttt{container.nu()}, \texttt{container.nw()}, \texttt{container.nq()}: it returns the total number of processes in CPU, WL and MQ
  \item \texttt{container.meanNU}, \texttt{container.meanNW}, \texttt{container.meanNQ}: global mean number of processes  in CPU, WL and MQ in the latest $\pi$ CS
\end{itemize}

\begin{itemize}
  \item \texttt{array}: for the sake of simplicity, this is the definition of a vector element of type \texttt{float}
  \item \texttt{array[s] X}: it is an array that stores the $s$ values chosen to decide the scheduler's state
  \item \texttt{getX(container cont)}: it returns the current state of the object \texttt{cont}
  \item \texttt{float slice}: value to be assigned to the next time slice
  \item \texttt{list Slist}: it's the list storing the latest $m$ values assigned to the time slice, except the current one
  \item \texttt{list Zlist}: it's the list storing the latest $m$ satte vectors, except the current one 
  \item \texttt{rcond(array[a][a] M)}, \texttt{inv(array[a][a] M)}: they are functions returning the the conditioning number and the inverse of a ``square'' array  \texttt{M}
  \item \texttt{trans(array[a][b] M)}: it's the function returning the transpose \texttt{array[b][a] W} of a bi-dimensional array \texttt{M}
  \item \texttt{vcat(array[a][b] M, array[c][b] N)}: it returns \texttt{array[a+c][b] K} formed by the vertical concatenation of \texttt{M} and \texttt{N}
  \item \texttt{vprod(array[a][b] M, array[b][c] N)}: vector product of two arrays
\end{itemize}

\begin{itemize}
  \item \texttt{list Elist}: a list of lists; each element\texttt{Elist(i)} contains a list of one-step prevision errors given by the $i$-th model in the latest $m$ CS
  \item \texttt{list ABlist}: each element of the list is an array $s\times s+1$ correspondent to the concatenation of the terms of the couple $(A_i,B_i)$ that describes the $i$-th model
  \item \texttt{int argmin(array[s] V)}:  it's a function returning the index corresponding to the first minimum in the vector \texttt{V}
\end{itemize}

\newpage

\subsection{ Sample code}

\noindent \textbf{Dispatcher's routine}

\bigskip

\small{
%\begin{center}
\begin{tabular}{l}
\verb|float tu, tw, tq|\\
\verb|list L = cont.getID //list's initialization |\\
\verb|int n = cont.getn() // number of processes |\\
\verb|proc p = new proc() // temporary marker's initialization |\\
\verb|for i=1 to n // active processes' cycle |\\
\verb|  p = cont.getProc(L(i))|\\
\verb|  update(p) // set updating operations of each process |\\
\verb|  tu = tu + p.U|\\
\verb|  tw = tw + p.W|\\
\verb|  tq = tq + p.Q|\\
\verb|end|\\
\verb|tu = tu/n // current CS mean calculation |\\
\verb|tw = tw/n|\\
\verb|tq = tq/n|\\
\verb|cont.Ulist.del(1) // elimination of the first element in the list |\\
\verb|cont.Wlist.del(1)|\\
\verb|cont.Qlist.del(1)|\\
\verb|cont.Ulist.append(tu) // updating of the last element |\\
\verb|cont.Wlist.append(tw)|\\
\verb|cont.Qlist.append(tq)|\\
\verb|cont.meanTU = mean(Ulist) // mean calculation |\\
\verb|cont.meanTW = mean(Wlist)|\\
\verb|cont.meanTQ = mean(Qlist)|\\
\verb|cont.NUlist.del(1)|\\
\verb|cont.NWlist.del(1)|\\
\verb|cont.NQlist.del(1)|\\
\verb|cont.NUlist.appen(cont.nu())|\\
\verb|cont.NWlist.appen(cont.nw())|\\
\verb|cont.NQlist.appen(cont.nq())|\\
\verb|cont.meanNU = mean(NUlist)|\\
\verb|cont.meanNW = mean(NWlist)|\\
\verb|cont.meanNQ = mean(NQlist)|\\
\end{tabular}
%\end{center}
}
\bigskip
\bigskip

\normalsize

\noindent \textbf{Academic Linear Regression}

\bigskip

\small{
\begin{tabular}{l}
\verb|array[s] X = getX(cont) // get the current state|\\
\verb|list oldZlist = new list(Zlist) // copy the list before any updating|\\
\verb|Zlist.del(1)  //eliminate the first element|\\
\verb|Zlist.append(X)  // update the tail of the list|\\
\verb|array[s][m] Z = oldZlist.getArray()|\\
\verb|array[m] S = Slist.getArray()|\\
\verb|array[s+1][m] V1 = vcat(Z,S)|\\
\verb|array[s][m] V2 = Zlist.getArray()|\\
\verb|array[s+1][s+1] M = vprod(V1,trans(V1))|\\
\verb|array[s][s+1] AB // initialize a proper array|\\
\verb|if rcond(M)<threshold // the threshold depends on the machine's precision|\\
\verb|  trow Exception e // exception handler|\\
\verb|else|\\
\verb|  AB = vprod(V2,vprod(trans(V1),inv(M)))|\\
\verb|  // AB is the matrix that define our model|\\
\verb|end|\\
\end{tabular}
}

\normalsize

\bigskip
\bigskip

\noindent \textbf{Switching Control routine}

\bigskip

\small{
\begin{tabular}{l}
\verb|array[r] Emean|\\
\verb|for i=1 to r|\\
\verb|  Elist(i).del(1)|\\
\verb|  float e|\\
\verb|  e = Zlist(m)-vprod(ABlist(i),vcat(Zlist(m-1),Slist(m-1)))|\\
\verb|  e = e*e|\\
\verb|  Elist(i).append(e)|\\
\verb|  Emean[i-1] = mean(Elist(i).getArray())|\\
\verb|end|\\
\verb|p = argmin(Emean)|\\
\verb|if p==1|\\
\verb|  timeslice = control_1|\\
\verb|else if p==2|\\
\verb|  timeslice = control_2|\\
\verb|  ...|\\
\verb|else if p==r|\\
\verb|  timeslice = control_r|\\
\verb|end|\\
\verb|Slist.del(1)|\\
\verb|Slist.append(u)|\\
\end{tabular}
}

\bigskip
\bigskip

\normalsize

\noindent \textbf{Adaptive Control routine}

\bigskip

\small{
\begin{tabular}{l}
\verb|array[s] B|\\
\verb|for i=0 to s-1|\\
\verb|  B[i] = AB[i][s+1]|\\
\verb|end|\\
\verb|float u = -vprod(trans(X),AB)/vprod(trans(B),B)|\\
\verb|if u < tmin|\\
\verb|  u = tmin|\\
\verb|else if u > tmax|\\
\verb|  u = tmax|\\
\verb|end|\\
\verb|timeslice = u|\\
\verb|Slist.del(1)|\\
\verb|Slist.append(u)|\\
\end{tabular}
}

\normalsize

\newpage

\bibliographystyle{unsrt}
\bibliography{biblio}

\begin{thebibliography}{10}

\bibitem{bib:book05}
L.~Breuer and D.~Baum.
\newblock {\em An Introduction to Queueing Theory and Matrix-Analytic Methods}.
\newblock Springer, 2005.

\bibitem{bib:book07}
Peter Brucker.
\newblock {\em Scheduling Algorithms}.
\newblock Springer, 2007.

\bibitem{bib:book08}
Michael~L. Pinedo.
\newblock {\em Scheduling: Theory, Algorithms and Systems}.
\newblock Springer, 2008.

\bibitem{bib:book09}
Kenneth~R. Baker and Dan Trietsch.
\newblock {\em Principles of Sequencing and Scheduling}.
\newblock Wiley, 2009.

\bibitem{bib:book06}
Jeffrey~W. Herrmann.
\newblock {\em Handbook of Production Scheduling}.
\newblock Springer, 2006.

\bibitem{bib:book75}
Leonard Kleinrock.
\newblock {\em Queuing Systems}.
\newblock Wiley, 1975.

\bibitem{bib:book93}
T.L.~Magnanti R.K.~Ahuja and J.B. Orlin.
\newblock {\em Network Flows}.
\newblock Prentice Hall, 1993.

\bibitem{bib:book00}
R.~German.
\newblock {\em Performance Analysis of Communication Systems with Non-Markovian
  Stochastic Petri Nets}.
\newblock John Wiley \& Sons, 2000.

\bibitem{bib:book05b}
Giorgio~C. Buttazzo.
\newblock {\em Hard Real-Time Computing Systems: Predictable Scheduling
  Algorithms and Applications}.
\newblock Springer, 2005.

\bibitem{laplante}
Philip~A. Laplante.
\newblock {\em Real-Time Systems Design and Analysis}.
\newblock Wiley - IEEE Press, 2004.

\bibitem{blaze}
Jacek Blazewicz, Klaus~H. Ecker, Erwin Pesch, G{\"u}nter Schmidt, and Jan
  Weglarz.
\newblock {\em Scheduling Computer and Manufacturing Processes}.
\newblock Springer, 2001.

\bibitem{oliver}
Oliver Sinnen.
\newblock {\em Task Scheduling for Parallel Systems}.
\newblock Wiley, 2007.

\bibitem{bib:RT91}
John~A. Stankovic and Krithi Ramamritham.
\newblock The spring kernel: A new paradigm for real-time systems.
\newblock {\em IEEE Software}, 8:62--72, 1991.

\bibitem{ssac}
S.~Saez and A.~Crespo.
\newblock Dynamic scheduling solutions for real-time multiprocessor systems.
\newblock {\em Control Engineering Practice}, 5(7):1007--1013, 1997.

\bibitem{mltijl}
Marin Litoiu, Traian~C. Ionescu, and Jesus Labarta.
\newblock Dynamic task scheduling in distributed real time systems using fuzzy
  rules.
\newblock {\em Microprocessors and Microsystems}, 21(5):299--311, 1998.

\bibitem{fusiello}
Alan~A. Bertossi and Andrea Fusiello.
\newblock Rate-monotonic scheduling for hard-real-time systems.
\newblock {\em European Journal of Operational Research}, 99(3):429--443, 1997.

\bibitem{effective}
Peter Cowling and Marcus Johansson.
\newblock Using real time information for effective dynamic scheduling.
\newblock {\em European Journal of Operational Research}, 139(2):230--244,
  2002.

\bibitem{palopoli}
Gabriele Bolognini Benedetto~Allotta Luigi~Palopoli, Luca~Abeni and Fabio
  Conticelli.
\newblock Novel scheduling policies in real-time multithread control system
  design.
\newblock {\em Control Engineering Practice}, 10(10):1091--1110, 2002.

\bibitem{libe}
Daniel Liberzon.
\newblock {\em Switching in Systems and Control}.
\newblock Birkhauser, 2003.

\bibitem{giarre}
L.~Giarr{\'e}, D.~Bauso, P.~Falugi, and B.~Bamieh.
\newblock Lpv model identification for gain scheduling control: An application
  to rotating stall and surge control problem.
\newblock {\em Control Engineering Practice}, 14(4):351--361, 2006.

\bibitem{rtc}
Johan Eker, Per Hagander, and Karl-Erik {\AA}rz{\'e}n.
\newblock A feedback scheduler for real-time controller tasks.
\newblock {\em Control Engineering Practice}, 8(12):1369--1378, 2000.

\bibitem{channel}
Henrik Rehbinder and Martin Sanfridson.
\newblock Scheduling of a limited communication channel for optimal control.
\newblock {\em Automatica}, 40(3):491--500, 2004.

\bibitem{bib:telec99}
V.~Bharghavan Lu~Songwu and R.~Srikant.
\newblock Fair scheduling in wireless packet networks.
\newblock {\em IEEE/ACM Transactions on Networking}, 7(4):473--489, 1999.

\bibitem{bib:telec05}
Francesco~Delli Priscoli and Alberto Isidori.
\newblock A control-engineering approach to integrated congestion control and
  scheduling in wireless local area networks.
\newblock {\em Control Engineering Practice}, 13(5):541--558, 2005.

\bibitem{bib:telec02}
Satish Damodaran and Krishna~M. Sivalingam.
\newblock Scheduling algorithms for multiple channel wireless local area
  networks.
\newblock {\em Computer Communications}, 25(14):1305--1314, 2002.

\bibitem{bib:robot94}
R.G. Simmons.
\newblock Structured control for autonomous robots.
\newblock {\em IEEE Transactions on Robotics and Automation}, 10(1):34--3,
  1994.

\bibitem{bib:robot05}
D.~Robert D.~Simon and O.~Sename.
\newblock Robust control/scheduling co-design: application to robot control.
\newblock In {\em 11th IEEE Real Time and Embedded Technology and Applications
  Symposium (RTAS '05)}, pages 118--127, 2005.

\bibitem{mmmh}
Maode Ma and Mounir Hamdi.
\newblock An adaptive scheduling algorithm for differentiated services on wdm
  optical networks.
\newblock {\em Computer Communications}, 27(9):857--867, 2004.

\bibitem{super}
Seong-Jin Parka and Jung-Min Yang.
\newblock Supervisory control for real-time scheduling of periodic and sporadic
  tasks with resource constraints.
\newblock {\em Automatica}, 45(11):2597--2604, 2009.

\bibitem{multi}
Tei-Wei~Kuo Shi-Wu~Lo and Kam-Yiu Lam.
\newblock Multi-disk scheduling for time-constrained requests in raid-0
  devices.
\newblock {\em Journal of Systems and Software}, 76(3):237--250, 2005.

\bibitem{handbook}
Jacek Blazewicz, Klaus~H. Ecker, Erwin Pesch, G.~Schmidt, and Jan Weglarz.
\newblock {\em Handbook on Scheduling: From Theory to Applications}.
\newblock Springer, 2007.

\bibitem{deadline}
John~A. Stankovic, Marco Spuri, Krithi Ramamritham, and Giorgio~C. Buttazzo.
\newblock {\em Deadline Scheduling for Real-Time Systems - EDF and Related
  Algorithms}.
\newblock Kluwer Academic Publishers, 1998.

\bibitem{kai}
T.~Kailath, Ali~H. Sayed, and Babak Hassibi.
\newblock {\em Linear Estimation}.
\newblock Prentice Hall, 2000.

\bibitem{mosca}
Edoardo Mosca.
\newblock {\em Optimal, Predictive and Adaptive Control}.
\newblock Prentice Hall, 1995.

\bibitem{Khalil}
Hassan~K. Khalil.
\newblock {\em Nonlinear Systems}.
\newblock Premtice Hall, 1996.

\bibitem{isi}
A.~Isidori.
\newblock {\em Nonlinear Control Systems}.
\newblock Springer, 2001.

\bibitem{joseph}
James~H. Anderson.
\newblock {\em Handbook of Scheduling; Algorithms, Models and Performance
  Analysis}.
\newblock Chapman \& Hall-CRC, 2004.

\end{thebibliography}
\end{document}